\documentclass{article}

\usepackage{microtype}
\usepackage{graphicx}
\usepackage{booktabs} 

\usepackage{hyperref}

\usepackage[accepted]{icml2025}

\usepackage{amsmath}
\usepackage{amssymb}
\usepackage{mathtools}
\usepackage{amsthm}

\usepackage[capitalize,noabbrev]{cleveref}

\theoremstyle{plain}
\newtheorem{theorem}{Theorem}[section]

\theoremstyle{definition}
\newtheorem{definition}[theorem]{Definition}

\theoremstyle{remark}

\usepackage[textsize=tiny]{todonotes}

\usepackage{tabularx}
\usepackage{varwidth}
\usepackage{colortbl}
\definecolor{mygray}{gray}{0.9}
\definecolor{deepgray}{gray}{.8}
\definecolor{myblue}{cmyk}{.3,0,0,0}
\definecolor{deepblue}{cmyk}{.3,0,0,0}
\definecolor{SoftRed}{RGB}{255,200,200}
\definecolor{SoftGreen}{RGB}{200,255,200}
\definecolor{SoftPurple}{RGB}{220,200,255}
\definecolor{SoftBlue}{RGB}{200,220,255}
\definecolor{lightpurple}{RGB}{230, 200, 255}
\definecolor{lightyellow}{RGB}{255, 255, 204}

\newcommand{\coloredbox}[2]{%
    \colorbox{#1}{\begin{varwidth}[t]{\dimexpr\linewidth-2\fboxsep\relax}#2\end{varwidth}}%
}

\usepackage{xspace}

\newcommand{\name}{\texttt{Cape}\xspace}
\newcommand{\bert}{\texttt{Cape} (BERT)\xspace}
\newcommand{\gpt}{\texttt{Cape} (GPT2)\xspace}

\usepackage{pifont}
\usepackage{tikz}
\newcommand*\emptycirc[1][0.8ex]{\tikz\draw (0,0) circle (#1);} 
\newcommand*\halfcirc[1][0.8ex]{%
	\begin{tikzpicture}
	\draw[fill] (0,0)-- (90:#1) arc (90:270:#1) -- cycle ;
	\draw (0,0) circle (#1);
	\end{tikzpicture}}
\newcommand*\fullcirc[1][0.8ex]{\tikz\fill (0,0) circle (#1);} 

\usepackage[linesnumbered,ruled,vlined, algo2e]{algorithm2e} 
\usepackage{xspace}
\usepackage{color}
\usepackage{booktabs}
\usepackage{amsmath,amsfonts,amssymb,stmaryrd,xspace,enumitem}
\usepackage{subcaption}
\usepackage{multicol}
\usepackage{multirow}
\usepackage{amsthm}
\usepackage{xcolor}
\usepackage{tablefootnote}

\icmltitlerunning{CAPE: Context-Aware Prompt Perturbation Mechanism with Differential Privacy}

\begin{document}

\twocolumn[
\icmltitle{Cape: \underline{C}ontext-\underline{A}ware \underline{P}rompt P\underline{e}rturbation Mechanism with Differential Privacy}




\begin{icmlauthorlist}
\icmlauthor{Haoqi Wu}{tiktok}
\icmlauthor{Wei Dai}{tiktok}
\icmlauthor{Li Wang}{tiktok}
\icmlauthor{Qiang Yan}{tiktok}
\end{icmlauthorlist}

\icmlaffiliation{tiktok}{TikTok}

\icmlcorrespondingauthor{Haoqi Wu}{haoqi.1997@tiktok.com}

\icmlkeywords{differential privacy, private selection, large language model, black-box inference}

\vskip 0.3in
]



\printAffiliationsAndNotice{}  

\begin{abstract} 
Large Language Models (LLMs) have gained significant popularity due to their remarkable capabilities in text understanding and generation. However, despite their widespread deployment in inference services such as ChatGPT, concerns about the potential leakage of sensitive user data have arisen. Existing solutions primarily rely on privacy-enhancing technologies to mitigate such risks, facing the trade-off among efficiency, privacy, and utility. To narrow this gap, we propose \name, a \underline{c}ontext-\underline{a}ware \underline{p}rompt p\underline{e}rturbation mechanism based on differential privacy, to enable efficient inference with an improved privacy-utility trade-off. Concretely, we introduce a hybrid utility function that better captures the token similarity. Additionally, we propose a bucketized sampling mechanism to handle large sampling space, which might lead to long-tail phenomenons. Extensive experiments across multiple datasets, along with ablation studies, demonstrate that \name achieves a better privacy-utility trade-off compared to prior state-of-the-art works.
\end{abstract}

\section{Introduction}\label{sec:intro}
\begin{figure}[h!]
    \setlength{\abovecaptionskip}{3pt}
    \setlength{\belowcaptionskip}{0pt}
    \centering
    \includegraphics[width=0.9\linewidth]{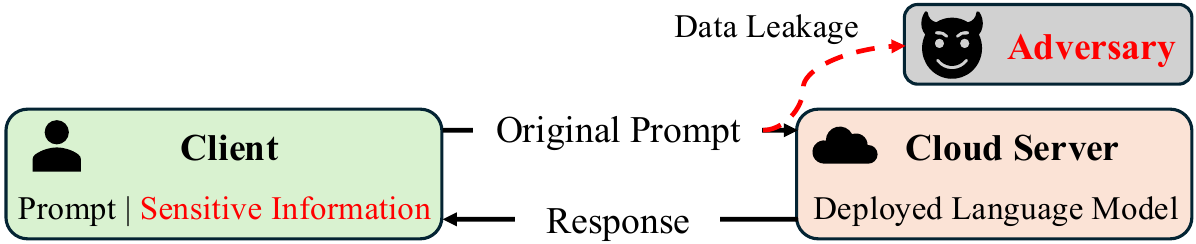}
    \caption{Illustration of existing LLM inference services. 
    }
    \label{fig:scenario}
\end{figure}

Recent advancements in large language models~\cite{attention-all-need-17} have revolutionized fields such as natural language processing~\cite{bert-19, gpt2-19}, driving their widespread adoption in machine learning (ML) inference services. However, despite their growing popularity, data privacy remains a critical concern.
In ML services like ChatGPT~\cite{chatgpt}, the server provides an API that processes client queries and returns generated responses. This paradigm (Figure~\ref{fig:scenario}) requires clients to transmit their queries—commonly referred to as prompts in the context of language models—to the server in plaintext, posing significant risks to user privacy, as the prompts can contain sensitive user information or confidential business guidelines~\cite{pp-instruction-icml24}.
For example, a user might request ChatGPT to summarize or analyze a confidential business report~\cite{samsung-leakage}, potentially resulting in severe privacy breaches. This issue is further exacerbated by the growing stringency of modern data privacy regulations.

Existing solutions~\cite{spu, puma-2023, ditto-24, ciphergpt-23, nimbus} utilize cryptographic techniques like secure multi-party computation (MPC)~\cite{yao-mpc} and homomorphic encryption (HE)~\cite{fhe-09} to offer provable security. However, the huge computation and communication overhead hinders their application in real world scenarios~\footnote{Take the SOTA 3-party computing work Ditto~\cite{ditto-24} as an example, the average runtime for generating one token with a sequence length of 128 on Bert-base model requires about 30s.}.
Another line of works utilize the client-server hybrid execution paradigm, where client performs some lightweight computations and sends shallow layer embeddings to the server. However, prior work~\cite{embedding-leakage} has shown that such information can be leveraged to reconstruct the original text effectively. Subsequent works~\cite{TextObfuscator-23, DP-Forward-23} bridge this gap by adding noise to intermediate embeddings to break the linkage between the noisy embeddings and original tokens. 
However, a major limitation is their reliance on a white-box inference setup, requiring shallow layers to be deployed on the client side, which demands extensive model modifications, making it seldom practical.

A promising approach is to utilize differential privacy (DP)~\cite{dp-06},
widely adopted as the \textit{de facto} standard for protecting user privacy, to verbatim perturb tokens in the prompts. It is highly efficient and requires no modifications to the back-end model.
SANTEXT~\cite{santext-21} and CUSTEXT~\cite{custext-22} focus on text classification tasks. 
InferDPT~\cite{privinfer-23} takes the first step towards the open-ended text generation tasks 
that manages to extract more accurate generation from the perturbed response from the server with the aid of a local LLM.   
To enhance utility, these works replace the original token with a random token from `truncated' adjacency lists, to reduce the large sampling space in NLP (e.g., BERT models have a vocabulary of size 30,522). However, this practice achieves higher utility at the sacrifice of privacy, making it vulnerable to $k$-nearest neighbor attacks~\cite{embedding-leakage}.
Besides, these works typically measure semantic similarity via token-level embedding distances.
Such approach overlooks contextual information, often leading to reduced contextual semantic coherence.

\begin{figure}[h]
    \setlength{\abovecaptionskip}{1pt}
    \setlength{\belowcaptionskip}{0pt}
    \centering
    \includegraphics[width=0.7\linewidth]{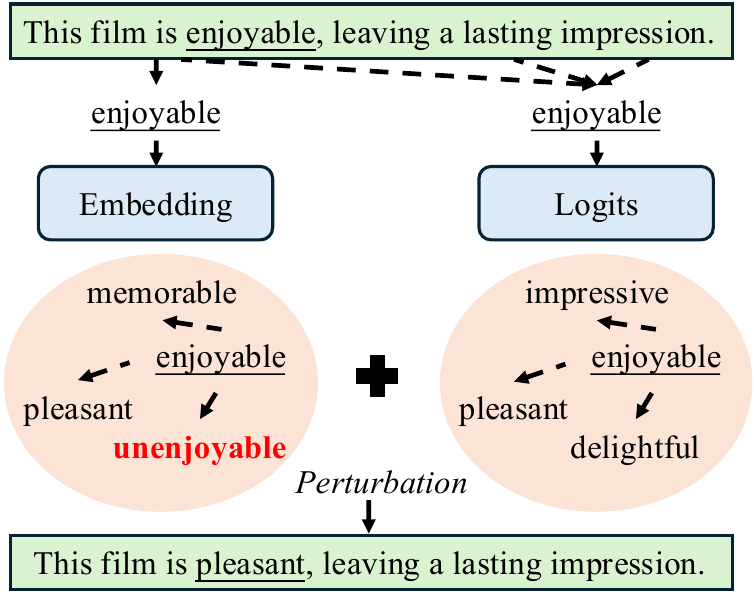}
    \caption{Intuition: Drawback of embedding distance alone.}
    \label{fig:hybrid-intuition}
\end{figure}

\textbf{A Motivating Example:} As shown in Figure~\ref{fig:hybrid-intuition}, `enjoyable' and `unenjoyable' appear close in embedding space despite conveying opposite meanings. Incorporating contextual information as a constraint can mitigate such ambiguities. 

Therefore, \textit{Can we perform efficient, context-aware prompt perturbation with better privacy-utility trade-off?}

As an answer to the above question, we develop \name, a \textit{context-aware} and \textit{bucketized} DP mechanism tailored to enhance semantic similarity measurement and optimize the sampling process in large sampling spaces for NLP tasks.

\begin{itemize}[itemsep=-0.3em, topsep=-2em]
    \item \textbf{Context-Aware Prompt Perturbation}. We introduce a hybrid utility function that integrates both \textit{contextual information} and \textit{token embedding distance} (e.g., Euclidean distance) that achieves improved utility. Such utility function is essential to the subsequent DP-based perturbation (formulated as \textit{private selection} problems), to ensure rigorous privacy guarantees.
    \item \textbf{Bucketized Sampling Mechanism}. We propose a customized exponential mechanism to tackle the challenge of large sampling spaces in NLP domain.
    Specifically, we utilize equal-width bucketing to restrict the sampling probability of low-utility items, achieving an improved privacy-utility trade-off. 
    \item \textbf{Empirical Evaluations.} We conduct extensive experiments on both text classification and text generation tasks across three datasets. The experiments confirm that \name achieves a better privacy-utility trade-off and provides stronger defense against existing privacy attacks. Additionally, ablation studies on the hyper-parameters in utility function and bucketized sampling further validate their effectiveness in our approach.
\end{itemize}

\section{Related Work}\label{sec:related-work}
Existing works typically employ privacy enhancing technologies to enable private inference, where exist a trade-off among efficiency, utility, and privacy.
Cryptographic-based solutions provide provable security and nearly lossless inference yet at the sacrifice of efficiency. The state-of-the-art works Ditto~\cite{ditto-24} and CipherGPT~\cite{ciphergpt-23} offers comparable accuracy to plaintext computation. However, they require at least 30s to infer a token on Bert-base model, which is not ready for practical scenarios.

\begin{table}[h]
    \setlength{\tabcolsep}{1.8pt}
    \setlength{\abovecaptionskip}{0pt}
    \setlength{\belowcaptionskip}{0pt}
    \centering
    \caption{Comparison of different methods. $\checkmark$ indicates the framework supports a feature, $\fullcirc$, $\halfcirc$ and $\emptycirc$ refer to high-, medium- and low-performance, respectively.}
    \scalebox{0.7}{
    \begin{tabular}{@{}c|ccccc@{}}
\toprule
\textbf{Method} & \textbf{Black-box}         & \textbf{Inference}         & \textbf{Privacy} & \textbf{Efficiency} & \textbf{Utility} \\ \midrule
CipherGPT~\cite{ciphergpt-23}       & $\checkmark$               & $\checkmark$               & $\fullcirc$      & $\emptycirc$        & $\fullcirc$      \\
Ditto~\cite{ditto-24}           & $\checkmark$               & $\checkmark$               & $\fullcirc$      & $\emptycirc$        & $\fullcirc$      \\ \midrule
TextObfuscator~\cite{TextObfuscator-23}  & \ding{55} & $\checkmark$               & $\halfcirc$      & $\halfcirc$         & $\halfcirc$      \\
DP-Forward~\cite{DP-Forward-23}      & \ding{55} & $\checkmark$               & $\halfcirc$      & $\halfcirc$         & $\halfcirc$      \\ \midrule
SANTEXT~\cite{santext-21}         & $\checkmark$               & \ding{55} & $\halfcirc$      & $\fullcirc$         & $\halfcirc$      \\
CUSTEXT~\cite{custext-22}         & $\checkmark$               & \ding{55} & $\halfcirc$      & $\fullcirc$         & $\halfcirc$      \\
InferDPT~\cite{privinfer-23}        & $\checkmark$               & $\checkmark$               & $\halfcirc$      & $\fullcirc$         & $\halfcirc$      \\ \midrule
$\name$         & $\checkmark$               & $\checkmark$               & $\halfcirc$      & $\fullcirc$         & $\halfcirc$      \\ \bottomrule
\end{tabular}
}
    \label{tab:work-comparison}
\end{table}


For DP-based solutions, although they yield better efficiency but faces the trade-off between privacy and utility. Some works add DP noise based on token embedding distance over the discrete vocabulary set. SANTEXT~\cite{santext-21} utilizes metric-LDP~\cite{metric-LDP-18} and provides privacy protection over the entire vocabulary. Subsequently, CUSTEXT~\cite{custext-22} and InferDPT~\cite{privinfer-23} prunes the sampling space to a `truncated' adjacency list (either static or dynamic), to achieve better utility at the sacrifice of privacy.
Furthermore, SANTEXT and CUSTEXT mainly focus on training data privacy and text classification tasks.
Our work introduces a context-aware utility function and optimized sampling mechanism tailored for large sampling space that strikes better trade-off.  
Other works like TextObfuscator~\cite{TextObfuscator-23} and DP-Forward~\cite{DP-Forward-23} adopt a client-server hybrid paradigm to offload some part of the model to the client. Consequently, they can add noise to the continuous intermediate embeddings, which shall incur smaller errors compared to sampling over a discrete domain. However, they require white-box access to the model and fine-tuning to align the embedding space, which could be hard to put into practice. In comparison, our work can be applied to black-box scenario, requiring minimal modifications to existing paradigm.
Recently, there are also some works that offer less privacy protection but potentially better utility for minimizing disclosure risks~\cite{reducing-privacy-acl24, llm-as-annoymizers-24}. These works use LLMs to detect and anonymize sensitive attributes in an innovative way, which we perceive as complementary to our approach for improving utility.

\section{Preliminaries}\label{sec:background}


\subsection{Transformer-based Inference}\label{sec:transformers}
Transformer models generally consist of three parts: 1) the \texttt{Embedding} module that maps discrete tokens to continuous word embeddings; 2) a stack of Transformer \texttt{Block}; 3) the last \texttt{Head} module that maps the hidden embeddings to task-specific \textit{logit}. 
The logits $y_t = \mathcal{M}(x_t|Ctx)$ are used to deduce the sampling probabilities for next token $x_t$, where $Ctx$ denotes the preceding text (and following text).

\subsection{Differential Privacy}\label{sec:dp} 
Differential privacy (DP) defines the \textit{worst-case} privacy guarantee. The maximum output probability difference between any two neighbor inputs is bounded by $e^{\epsilon}$, where $\epsilon$ denotes the privacy budget. 
In this paper, we focus on \textit{local} DP~\cite{ldp-08}, which operates on individual data points $x$ and allows clients to locally perturb their data before sending it to the server using the DP mechanism $\mathcal{R}$.
Therefore, an adversary cannot effectively infer the original input $x$ based on the outputs.

\begin{definition}[$\epsilon$-Local Differential Privacy~\cite{ldp-08}]
    Given a privacy budget $\epsilon > 0$, a randomized algorithm $\mathcal{R}: \mathcal{X} \rightarrow \mathcal{Y}$ satisfies $\epsilon$-LDP if, for each possible output $y \in \mathcal{Y}$, and all adjacent  $x_1, x_2 \in \mathcal{X}$, the following holds: 
    \begin{equation}
    \setlength{\abovedisplayskip}{0pt} 
    \setlength{\belowdisplayskip}{0pt} 
        \frac{\mathbb{P}[\mathcal{R}(x_1)=y]}{\mathbb{P}[\mathcal{R}(x_2)=y]} \leq e^{\epsilon}
    \end{equation}
\end{definition}


In the NLP domain, we tokenize the prompt into tokens, and the sampling spaces $\mathcal{X}$ and $\mathcal{Y}$ are defined as the token vocabulary $\mathcal{V}$. Conceptually, under $\epsilon$-LDP, any adjacent tokens $x_1$ and $x_2$ can be randomized to the same output token $y$ with indistinguishable probability.

\subsection{Private Selection}
The task of private selection is to design a randomized algorithm $\mathcal{R}$ that returns some $y_i$ that approximately maximizes the utility $u(x, y_i)$ while satisfying differential privacy.

\begin{definition}[$\epsilon$-Exponential Mechanism (EM)~\cite{em-07}]
    Given a privacy budget $\epsilon > 0$, a randomized Exponential mechanism $\mathcal{R}_{EM}: \mathcal{X} \rightarrow \mathcal{Y}$ satisfies $\epsilon$-LDP if, for each possible output $y_i \in \mathcal{Y}$, and input $x \in \mathcal{X}$:
    \begin{equation}
        \setlength{\abovedisplayskip}{0pt} 
        \setlength{\belowdisplayskip}{0pt} 
        \mathbb{P}[y_i|x] \propto \frac{\epsilon}{2\triangle} \cdot u(x, y_i)
    \end{equation}
    where $u(x, y_i)$ denotes the utility function that measures the utility score for selecting $y_i$, and $\triangle$ denotes the sensitivity, which is computed as $\triangle_u = \underset{x, x', y}{\mathrm{max}}|u(x, y) - u(x', y)|$.

\end{definition}

\subsection{Threat Model}\label{sec:threat}
In our scenario, the client $\mathcal{C}$ sends the prompt $x$ to cloud server $\mathcal{S}$ and receives generation results in return: $r \leftarrow \mathcal{M}(x)$. Here, $x = \{t_1, t_2, ..., t_n\}$, where $t_i \in \mathcal{V}$ denotes the tokens in the vocabulary. The prompt $x$ may contain sensitive tokens such as emails and genders, and directly uploading such information to $\mathcal{S}$ compromises the privacy of clients.
Table~\ref{tab:notations} lists the main used notations in this work. 

\begin{table}[h]
    \setlength{\abovecaptionskip}{0pt}
    \setlength{\belowcaptionskip}{0pt}
    \centering
    \caption{Notations used in this paper.}
    \scalebox{0.7}{
\begin{tabular}{@{}c|l@{}}
\toprule
\textbf{Symbol}        & \multicolumn{1}{c}{\textbf{Description}}                          \\ \midrule
$\mathcal{S}$          & the server that provides inference service                        \\
$\mathcal{C}$          & the client that inputs prompt                                     \\
$\mathcal{M}$          & the language model                                                \\
$\mathcal{V}$          & the vocabulary                                                    \\
$t$                    & word token $t \in \mathcal{V}$                                    \\
$x, \hat{x}$           & the original prompt $x = \{t_1, t_2,...,t_{n}\}$ and perturbed prompt \\ \midrule
$\epsilon$, $\triangle$             & privacy budget and sensitivity, respectively                                                    \\
$u(\cdot, \cdot)$      & utility function                                                  \\
$\mathcal{R}$          & the private selection mechanism                                   \\ \midrule
$d_{euc}(t, t')$       & Euclidean distance between tokens $t$ and $t'$                    \\
$L_r$                  & the logits for candidate token $t_r$ given context $L_r$          \\
$D(t_i, t_r)$          & the distance between $i$-th token $t_i$ and candidate token $t_r$ \\
$N_b$                    & the number of buckets                                                  \\
$\lambda_L, \lambda_D$ & importance factors of contextual logits and token distance        \\ \bottomrule
\end{tabular}
}
    \label{tab:notations}
\end{table}

\textbf{Goal.} Our goal is to perturb the original prompt using DP as $\hat{x} = \mathcal{R}(x)$, ensuring that not only the perturbed $\hat{x}$ provides \textit{rigorous privacy guarantee} (i.e., an adversary can not effectively infer original prompt) but also maintains an \textit{acceptable utility} (i.e., task-specific performance is retained). 

\textbf{Adversary.} We consider the LLM inference service provider $\mathcal{S}$ as the potential adversary. We do not consider man-in-the-middle attacks and assume the communication channel is secure.
$\mathcal{S}$ is assumed to be \textit{semi-honest}, meaning the adversary follows the execution flow for inference but may attempt to extract sensitive information by collecting and analyzing the messages received from the client. $\mathcal{S}$ has access to the perturbed prompt $\hat{x}$, and the DP mechanism $\mathcal{R}$ is publicly known. Moreover, $\mathcal{S}$ possesses unlimited computational power, allowing it to launch attacks, such as the mask token inference attack~\cite{santext-21}.

\section{Design}\label{sec:design}

\subsection{Context-aware Prompt Perturbation}\label{sec:workflow} 
\name acts as a safeguard layer, allowing the client to locally perturb the prompts and hide sensitive data from the server. 

\begin{figure}[h]
    \centering
    \setlength{\abovecaptionskip}{6pt}
    \setlength{\belowcaptionskip}{0pt}
    \includegraphics[width=\linewidth]{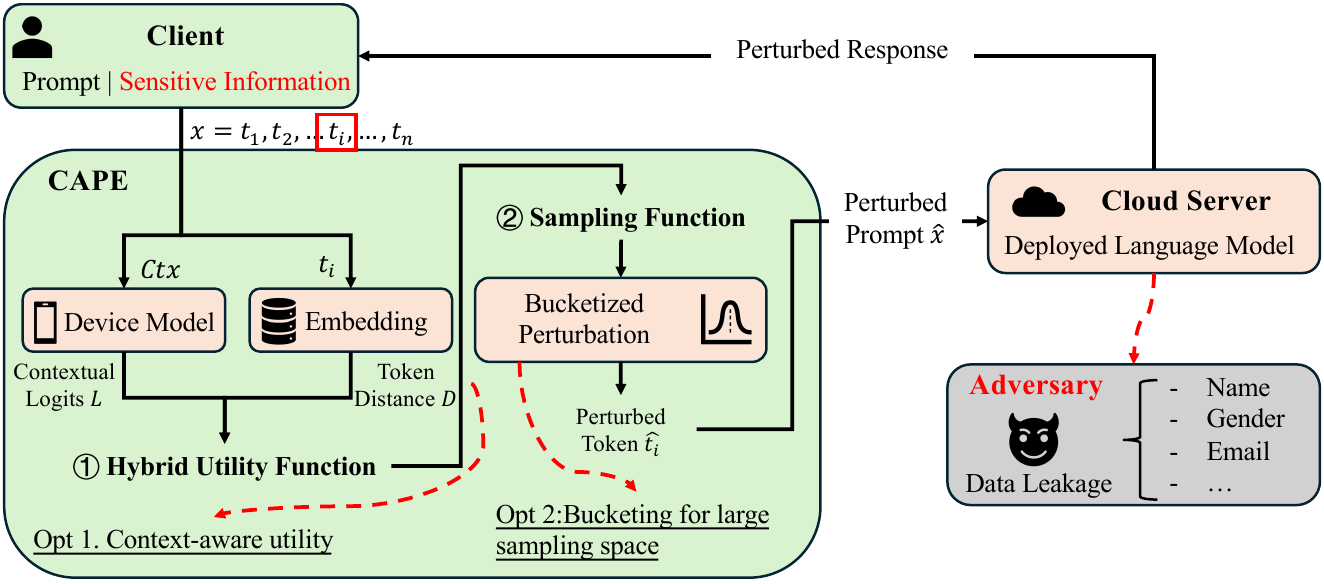}
    \caption{High-level workflow of \name.}
    \label{fig:workflow}
\end{figure}

As illustrated in Figure~\ref{fig:workflow} (in green background), the overall procedure is as follows: 
Given an input prompt containing a sequence of sensitive tokens $x=\{t_1, t_2, ..., t_n\}$, and a differential private mechanism $\mathcal{R}$ (i.e., \name), we verbatim replace the original token $t_i$ to a randomized output token $\hat{t}_i = t_r \in \mathcal{V}$. All these randomized tokens $\hat{t}_i$ constitute the perturbed prompt $\hat{x} = \{\hat{t}_i = \mathcal{R}(t_i)\}$, where $i \in [1, n]$.
Then, the perturbed prompt $\hat{x}$ is sent to $\mathcal{S}$ for subsequent text generation.
\name consists of two major components:
\begin{itemize}[itemsep=0em, topsep=0em]
    \item \textbf{Hybrid Utility Function:} By hybrid, we consider both token distance $D$ and contextual logits $L$ when computing the utility scores for candidate sampling tokens. Prior works~\cite{santext-21, custext-22} merely focus on the token distance alone. We resort to a client-side small device model to capture the contextual information and use Euclidean distance as the default measurement for token distance. 
    \item \textbf{Bucketized Sampling Function:}
    We adopt Exponential mechanism, the \textit{de facto} standard in private selection problems, to sample replacement tokens.
    Notably, the large sampling space in NLP (e.g., the vocabulary in Bert-base models is 30,522) could lead to long-tail phenomenon. To tackle this problem, we propose a bucketized variant to refine the sampling probability distribution, especially when the budget $\epsilon$ is small.
\end{itemize}

\subsubsection{Utility Function}\label{sec:utility}
In private selection, the utility function is essential to the sampling performance, which should satisfy two features: 1) \textbf{Monotonicity:} Candidates with higher utility scores are more likely to be sampled, which guarantees the utility of the mechanism; 2) \textbf{Boundedness:} The output space of the utility function should be bounded to keep the sensitivity $\triangle$ controllable, ensuring rigorous privacy protection.

We hereby design a hybrid utility function for the $i$-th token $t_i$ in a prompt as: 
\begin{equation}\label{eq:utility}
    \setlength{\abovedisplayskip}{3pt} 
    \setlength{\belowdisplayskip}{3pt} 
    u(t_i, t_r) = L_r^{\lambda_L} \cdot D(t_i, t_r)^{\lambda_D}
\end{equation}
where $t_i, t_r$ denote the original and candidate tokens, $L_r = \mathcal{M}_c(t_r | Ctx)$ denotes the logits for $t_r$ given context $Ctx$ and a client-side model $\mathcal{M}_c$. $Ctx$ can be both preceding and following information (e.g., $\{t_1,...,t_{i-1},t_{i+1},...,t_n\}$ in Bert) or  preceding information alone (e.g., $\{t_1, t_2,...,t_{i-1}\}$ in GPT). $D(t_i, t_r)$ denotes the token embedding distance between $t_i$ and $t_r$. 
Additionally, we introduce importance factors $\lambda_L$ and $\lambda_D$ for logit and distance, respectively.
Theoretically, increasing $\lambda_L$ enhances \textit{contextual coherence}, while emphasizing $\lambda_D$ improves \textit{semantic similarity}.


    




    


To satisfy the monotonicity, since a larger logit indicates higher contextual relevance (i.e., positive correlation) while a larger distance indicates lower semantic similarity (i.e., negative correlation), we use the form $u \propto L \cdot \mathbf{exp}(-D)$ to obtain the utility. By default, we use Euclidean distance over the embedding space to measure the token distance $D$. The contribution of token distance can be formulated as:
{\setlength{\jot}{-1px}
\begin{align}
    \setlength{\abovedisplayskip}{0pt} 
    \setlength{\belowdisplayskip}{0pt} 
    \Vec{D}_{raw} &= \{d_{euc}(t_i, t_r), 1 \leq r \leq |\mathcal{V}|\} \\[0px]
    \Vec{D}_{norm} &= \frac{\Vec{D} - D_{min}}{D_{max} - D_{min}} \\[0px]
       \Vec{D} &= \mathbf{exp}(-\Vec{D}_{norm})
\end{align}

Regarding the boundedness, since $\Vec{D}_{norm} \in [0, 1]$, we have $\Vec{D} \leq 1$. However, the logit $L$, which distributes over real value space $\mathbb{R}$, is dependent on the model and input. To limit the contribution of logit to the utility score, we can clip the logit to a given bound $B$ as follows:
\begin{equation}
    \setlength{\abovedisplayskip}{2pt} 
    \setlength{\belowdisplayskip}{2pt} 
    \Vec{L} = \mathbf{clip}(\Vec{L}_{raw}, B) 
\end{equation}
To determine a suitable bound $B$, we analyze the logit distribution using a few calibration samples. The bound is then selected based on the maximum value. 
We provide a histogram of both logits and distance in Appendix~\ref{append:hist}.



\subsubsection{Sampling Algorithm}\label{sec:sampling}

With the utility function defined, we proceed to the sampling algorithm, with Exponential mechanism (EM) as the basis.

\textbf{Long-Tail Phenomenon.}
We first note that there exists long-tail phenomenon when the sampling space is extremely large (i.e., a large vocabulary $\mathcal{V}$ in NLP). 
In standard EM, the sampling probability for $t_r$ can be simply formulated as:
\begin{equation}
    \mathbb{P}[\mathcal{R}(t_i) = t_r] = \frac{\mathrm{exp}(\frac{\epsilon}{2\triangle}u(t_i, t_r))}{\sum_{j=1}^{|\mathcal{V}|}\mathrm{exp}(\frac{\epsilon}{2\triangle}u(t_i, t_j))}
\end{equation}
When $\mathcal{V}$ is relatively large, many low-utility candidate tokens have individually negligible probabilities, but collectively their sum is significant.
According to the cumulative sampling probability distribution in Figure~\ref{fig:raw-cdf} and \ref{fig:candidate-cdf}, the majority of probabilities are significantly lower than 0.0001, yet they account for a substantial portion of the CDF. 
\begin{figure*}[h]
    \centering
    \setlength{\abovecaptionskip}{3pt}
    \setlength{\belowcaptionskip}{0pt}
    \begin{subfigure}[b]{0.3\linewidth}
        \includegraphics[width=\linewidth]{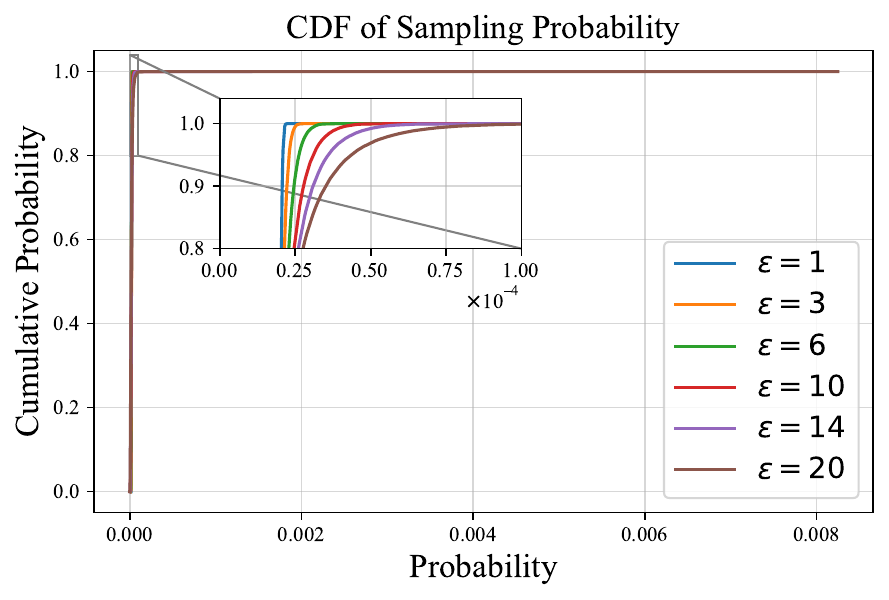}
        \caption{Sampling CDF.}
        \label{fig:raw-cdf}
    \end{subfigure}
    \hfill
    \begin{subfigure}[b]{0.3\linewidth}
        \includegraphics[width=\linewidth]{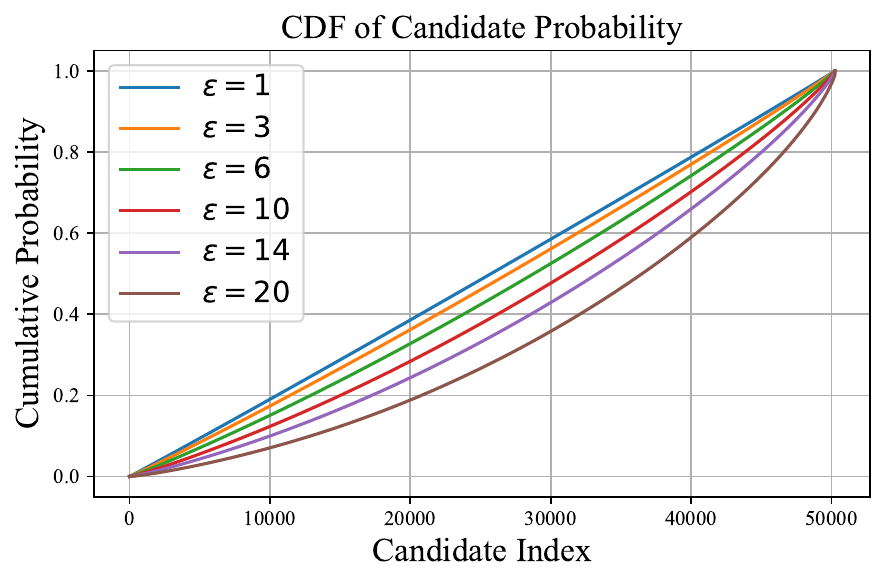}
        \caption{Candidate Probability.}
        \label{fig:candidate-cdf}
    \end{subfigure}
    \hfill
    \begin{subfigure}[b]{0.3\linewidth}
        \includegraphics[width=\linewidth]{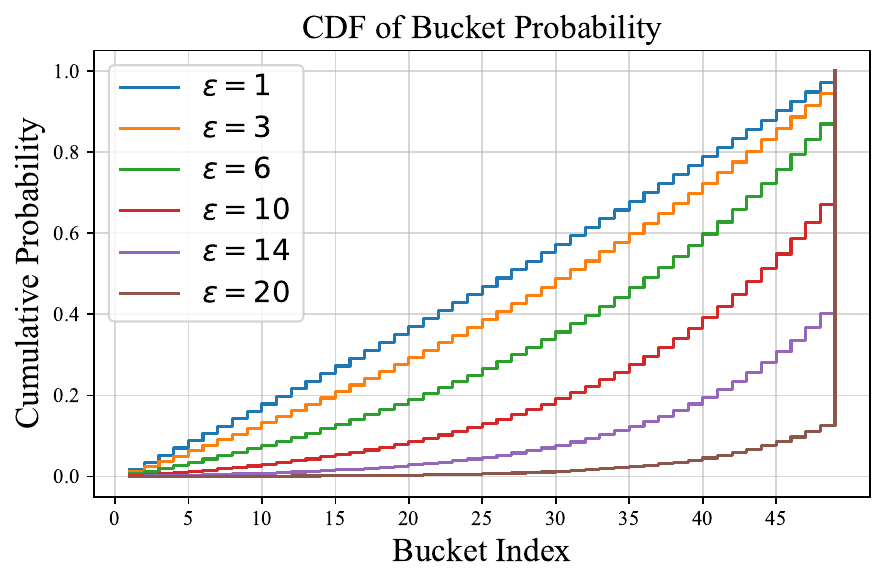}
        \caption{Bucket Probability.}
        \label{fig:bucket-cdf}
    \end{subfigure}
    \caption{Cumulative distribution function of sampling probabilities for `book' in `This is a good \underline{book}.' upon GPT2.}
    \label{fig:cdf}
\end{figure*}

Theoretically, let $p_s$ and $p_l$ denote the smallest and largest probabilities, respectively, with their maximum difference bounded by $e^{\epsilon}$ (i.e., $p_l \leq p_s \cdot e^{\epsilon}$). $N$ denotes the vocabulary size. The probability sum of the top-$k$ tokens ($\mathcal{K}$) relative to the remaining ones can be expressed as:
\begin{align}
    \setlength{\abovedisplayskip}{0pt} 
    \setlength{\belowdisplayskip}{0pt} 
    P_1 &= \sum_{i \in \mathcal{K}} p_i \leq k\cdot p_l,
    P_2 = \sum_{i \notin \mathcal{K}} p_i \geq (N-k)\cdot p_s \\
    \frac{P_1}{P_2} &\leq \frac{k\cdot p_l}{(N-k)\cdot p_s} 
    \leq \frac{k\cdot e^{\epsilon}}{(N-k)} \approx \frac{k\cdot e^{\epsilon}}{N}
\end{align}
If we set $\epsilon = 6$, $k=10$ and $N = 50000$, we have $\frac{P_1}{P_2} \leq \frac{403 \cdot 10}{50000}\approx 0.08$. That is, a great portion of low-utility candidates takes over $90\%$ percent of probability mass. In practice, the empirical cumulative probability for $k=10$ is less than $1\%$, i.e., $P_2 \ge 99\%$. 
This observation contradicts the intuitive expectation that candidate tokens closely resembling the original token would have higher probabilities.

\begin{algorithm}
\caption{Equal width Bucketing:  $\mathbf{Bucket}$}\label{algo:bucket}

    \KwIn{Utility score vector $\Vec{u}$, the number of buckets $N_b$}
    \KwOut{A set of buckets $B$ with different tokens.}

    Initialize $B \leftarrow \emptyset$
    
    $\Vec{u}_s = \mathbf{Sort}(\Vec{u})$

    $\Vec{u2b} = \mathbf{Bucketize}(\Vec{u}_s, N_b)$

    \For{$i \gets 1$ \KwTo $N_b$}{
        $b_i \leftarrow \{t_j| \Vec{u2b}_j==i\}$
        
        $\Vec{u}_{b,i} \leftarrow \Vec{u}\{b_i\}$

        \uIf{$\mathrm{len}(v) > 0$}{
            $B \leftarrow B | (\mathbf{mean}(\vec{u}_{b,i}), b_i)$
        }
    }
    \KwRet{B}
\end{algorithm}

\textbf{Bucketing Strategy.}
A straightforward solution is to use a larger $\epsilon$, which, however, results in a low level of formal privacy. To address this, we propose to bucketize the sampling space to refine the sampling probability distribution. Specifically, we adopt equal-width bucketing, which partitions the sampling space into uniformly sized $N_b$ buckets based on the numerical utility ranges.
The key idea is to bound the probability sum of all candidates in the same bucket (i.e., within a certain utility interval), proportional to $\mathrm{exp}(\frac{\epsilon}{2\triangle}\mathbf{mean}(b_i))$, where $b_i$ denotes the token set that maps to the $i$-th bucket, $\Vec{u}_{b,i}$ denotes the score vectors for these tokens and $\mathbf{mean}(b_i) = \sum_{k=1}^{|b_i|}b_{i, k} / |b_i|$. 
As shown in Algorithm~\ref{algo:bucket}, we sort the utility scores and assign each candidate to a unique bucket of width $(u_{max} - u_{min}) / N_b$. Notably, if a bucket contains no candidates, it is skipped-a scenario that may occur when $N_b$ is large. For each bucket, we use the average utility score of its candidates as its representative score.
Then, as shown in Algorithm~\ref{algo:cape-mechanism-em}, we use EM to sample a bucket, followed by a uniform sampling from the chosen bucket to obtain the replacement token.
The bucketized probability is given by:
\begin{equation}\label{eq:bucket-prob}
    \mathbb{P}[\mathcal{R}(t) = t_r] \propto \frac{\mathrm{exp}(\frac{\epsilon}{2\triangle}\mathbf{mean}(b_i))}{|b_i|},\ \mathrm{if}\ t_r \in b_i
\end{equation}
where $\triangle = \mathrm{max}_{t_r}\underset{b\sim b'}{\mathrm{max}}|b - b'|$, $b, b'$ denote utilities of different buckets.
As shown in Figure~\ref{fig:bucket-cdf}, \name effectively mitigates the long-tail phenomenon,
where low-utility candidates now constitute a smaller cumulative probability. 
The proof to Theorem~\ref{theo:bem} is provided in Appendix~\ref{append:proof}.

\begin{theorem}\label{theo:bem}
    The bucketized Exponential Mechanism satisfies $(\epsilon + \epsilon')$-differential privacy, where $\epsilon' = \ln(\underset{i, j}{\mathrm{max}}\frac{|b_i|}{|b_j|})$.
\end{theorem}

\subsection{Define Non-sensitive Tokens}

To further enhance the utility, we utilize a pre-defined set of context-independent non-sensitive tokens. This set primarily consists of 179 NLTK stopwords (e.g., `the', `a') and 32 punctuations (e.g., `,', `.'), which, while lacking significant context-sensitive information, are essential for text coherence and readability.
We retain these non-sensitive tokens while perturbing those sensitive ones in the prompt.

\setcounter{AlgoLine}{0}
\begin{algorithm}
\caption{\name Mechanism: $\mathcal{R}$~\label{algo:cape-mechanism-em}}

    \KwIn{Prompt $x=\{t_1, t_2, ..., t_n\}$, device model $\mathcal{M}$, embedding table $\Vec{e}$, importance factors $\lambda_L, \lambda_D$, bucket number $N_b$, and budget $\epsilon > 0$}
    \KwOut{Perturbed prompt $\hat{x} \leftarrow \mathcal{R}(x)$}

    Initialize $\hat{x} \leftarrow \emptyset$
    
    \For{$i \gets 1$ \KwTo $n$}{
        $\Vec{u}(t_i) = \mathbf{Utility}(\mathcal{M}, Ctx, \Vec{e}, \lambda_L, \lambda_D)$

        $\Vec{u}_b(t_i), \Vec{b}(t_i) \leftarrow \mathbf{Bucket}(\Vec{u}(t_i), N_b)$
        

        \tcc{Sample a bucket using EM}
        $b_r \leftarrow \mathbf{EM}(\Vec{u}_b(t_i), \Vec{b}(t_i), \epsilon)$
        
    
        \tcc{Randomly choose a token from $b_r$}
        $t_r \sim \mathbf{Uniform}(b_r)$

        $\hat{x} \leftarrow \hat{x} | t_r$
    }

    \KwRet{$\hat{x}$}
\end{algorithm}

\section{Experiments}\label{sec:experiments}

\textbf{Datasets \& Metrics.}
We consider two types of text tasks: 1) text classification; and 2) open-ended text generation.
For the former, we follow prior works~\cite{santext-21, custext-22} to use two GLUE datasets with privacy implications. 1) \textbf{SST-2:} This contains sentiment annotations for movie reviews, which is used to perform sentiment classification (positive or negative); 2) \textbf{QNLI:} This is a dataset containing sentence pairs for binary classification (entailment/not entailment). We use \textit{accuracy} as the metric.

For the latter, we follow~\cite{privinfer-23} to use \textbf{Wikitext-103-v1}, a large-scale dataset derived from Wikipedia articles for language modeling tasks. We use \textit{coherence}, and \textit{alignment} as the evaluation metrics (detailed in Section~\ref{sec:exp-open-gen}).

For privacy evaluation, we consider the following two attacks. We calculate the attack success rate $asr$ on sensitive tokens and the privacy score as $1 - asr$.

\textbf{KNN Attack~\cite{embedding-leakage}.} 
The adversary computes the embedding distance between the perturbed token and all other tokens in the vocabulary, and then selects the Top-$k$ tokens with the smallest distances, where we set $k=10$. The \textbf{Top-10 accuracy} is defined as the proportion of cases where the original token appears within Top-10.

\textbf{Masked Token Inference (MTI) Attack~\cite{santext-21}.} 
The adversary employs a BERT model, which captures both preceding and following contextual information, to predict the masked token. We replicate this attack by sequentially replacing the perturbed token $\hat{t}$ with the special token \textsf{[MASK]}. The \textbf{Rouge-L F1 score} is used to evaluate the similarity between predicted and original prompt.


\textbf{Baselines.} 
We compare \name with SANTEXT~\cite{santext-21} and CUSTEXT~\cite{custext-22} on their default setting (based on GloVe embedding). We also adopt InferDPT~\cite{privinfer-23} as the baseline (based on text-embedding-ada-002~\cite{ada-embedding}), which is the first work that supports open-ended text generation.
For \name, we utilize either the BERT~\cite{bert-19} model or the GPT-2~\cite{gpt2-19} model to capture contextual information, along with their respective embeddings to compute token distances. By default, we set $\lambda_L=0.5$, $\lambda_D=1.0$ and $N_b=50$.
We run inference on the original data as non-private baseline. 
The detailed vocabulary configuration is provided in Appendix~\ref{append:vocab}.
To ensure a fair comparison, we faithfully implement all these mechanisms in Python and re-run all the experiments.

\textbf{Experimental setup.}
All the experiments are carried out on one Debian 11 machine equipped with one Intel Xeon Platinum 8260 CPU (6 cores and 2.40GHz), 16GB of RAM and 4 Nvidia Tesla-V100-SXM2-32GB GPUs.

\subsection{Privacy-Utility Trade-off}\label{sec:tradeoff}
To begin, we focus on the \textbf{empirical} privacy-utility trade-off of different perturbation mechanisms in both text classification and open-ended text generation tasks. 
We will study the effects of \textbf{formal} privacy $\epsilon$ in the following Section~\ref{sec:influence-budget}.

In terms of efficiency, after a setup phase that computes the token embedding distance, the perturbation of \name only consumes about $0.1\sim0.15$ second for each input in average considering a small batch size of 2. The detailed efficiency evaluation is provided in Appendix~\ref{append:efficiency}.

\subsubsection{Text Classification}
For text classification tasks, we leverage a pre-trained Qwen2-1.5B-Instruct model~\cite{li2023towards} for zero-shot learning on the validation set of SST-2 and QNLI datasets. 
The instructions (i.e., system prompts) used to obtain classification results are provided in Appendix~\ref{append:prompt}. 
Notably, this inference-only paradigm presents a greater challenge in effectively preserving semantics, as the absence of re-training limits its ability to capture and refine noisy semantics.

Figure~\ref{fig:privacy-utility-knn-mti} illustrates the privacy-utility trade-off under KNN and MTI attacks with $\epsilon \in [1, 20]$, respectively.
The closer the line is to the \textbf{upper right} corner, the better the trade-off.
The detailed utility and privacy numbers are deferred to Appendix~\ref{append:classification-utility} and \ref{append:defense-score}.
The results across different settings confirm that \name achieves better trade-off than baselines.
\name achieves comparable task utility to CUSTEXT and superior to that of SANTEXT and InferDPT owing to the hybrid utility function and bucketing optimization. Specifically, CUSTEXT achieves higher utility even when $\epsilon$ is small (a stringent formal privacy) owing to its usage of fixed and small adjacency lists. However, such practice leads to worse defense against attacks.
Besides, \bert yields better utility than \gpt since Bert models captures both preceding and following contextual information, thus ensuring better coherence and semantic similarity.

\begin{figure}[h]
    \centering
    \setlength{\abovecaptionskip}{3pt}
    \setlength{\belowcaptionskip}{0pt}
    \begin{subfigure}[b]{0.49\linewidth}
        \includegraphics[width=\linewidth]{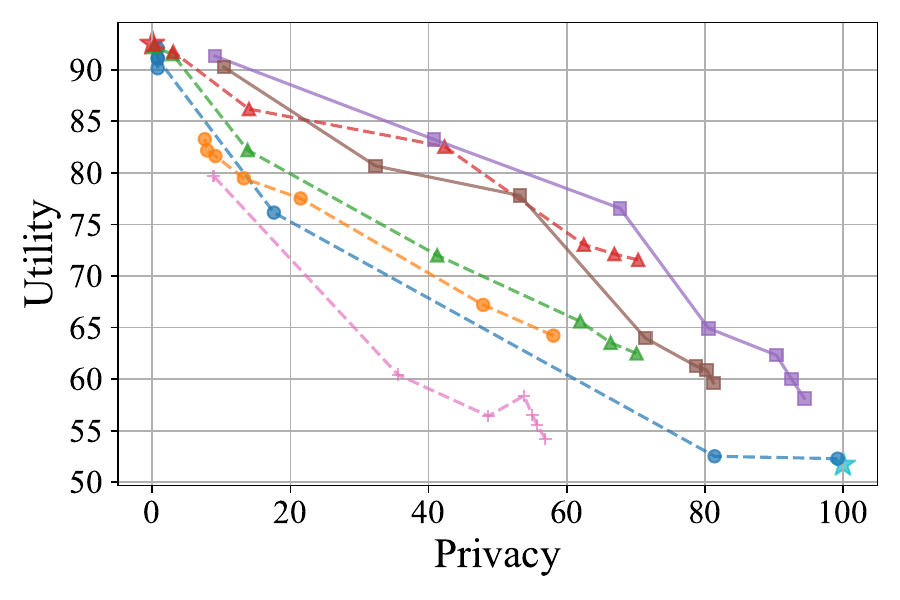}
        \caption{KNN attack on SST-2}
        \label{fig:trade-off-sst2-knn}
    \end{subfigure}
    \hfill
    \begin{subfigure}[b]{0.49\linewidth}
        \includegraphics[width=\linewidth]{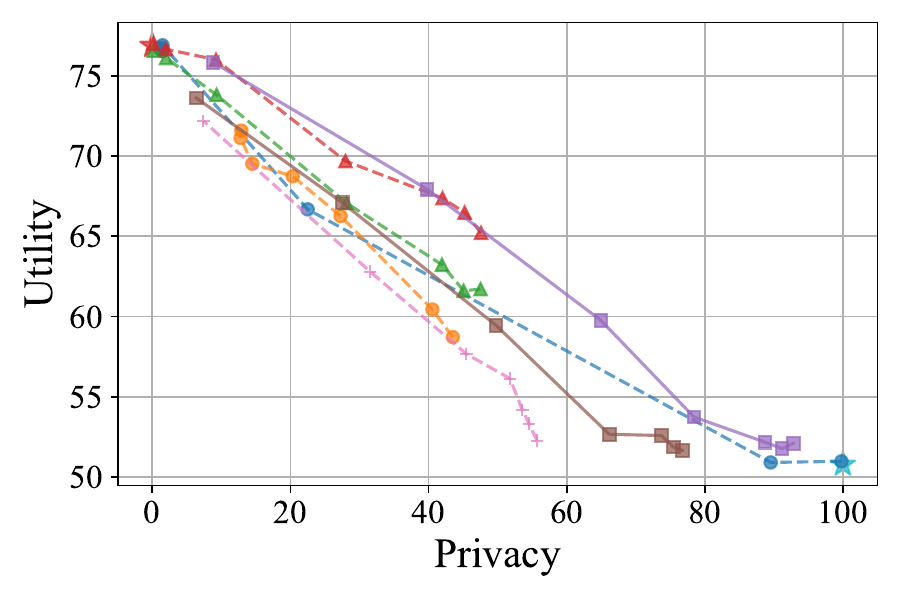}
        \caption{KNN attack on QNLI}
        \label{fig:trade-off-qnli-knn}
    \end{subfigure}
    \begin{subfigure}[b]{0.49\linewidth}
        \includegraphics[width=\linewidth]{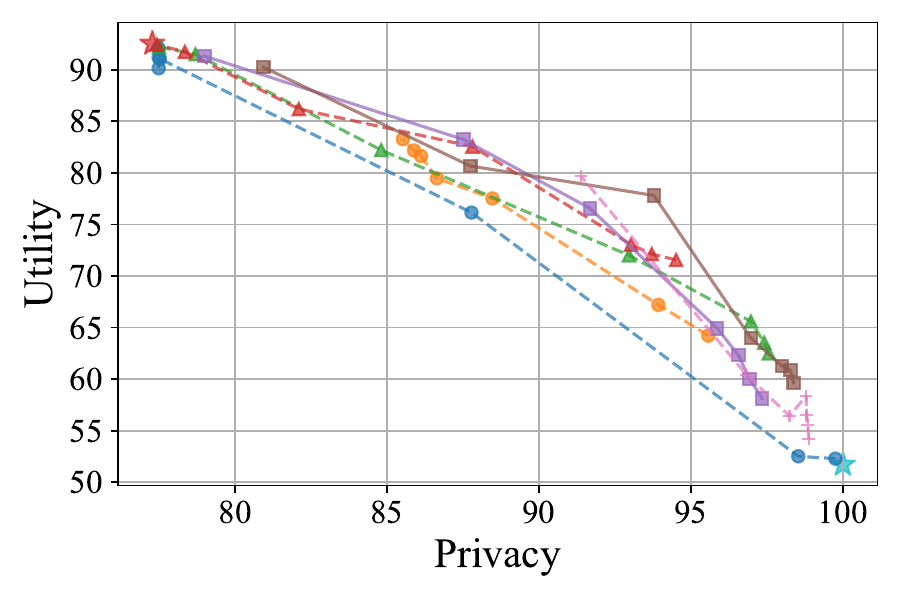}
        \caption{MTI attack on SST-2}
        \label{fig:trade-off-sst2-mt}
    \end{subfigure}
    \hfill
    \begin{subfigure}[b]{0.49\linewidth}
        \includegraphics[width=\linewidth]{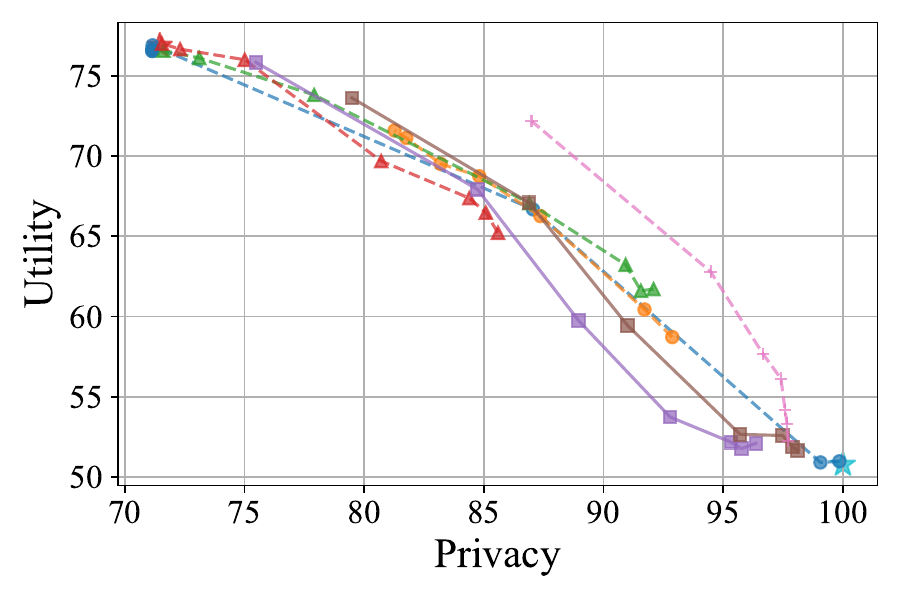}
        \caption{MTI attack on QNLI}
        \label{fig:trade-off-qnli-mt}
    \end{subfigure}
    \includegraphics[width=\linewidth]{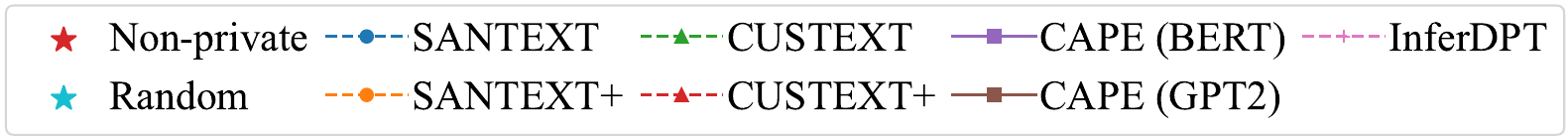}
    \caption{Privacy-utility trade-offs in terms of privacy attacks vs. accuracy rates by varying $\epsilon \in [1, 20]$. Privacy is measured by privacy scores under empirical attacks.}
    \label{fig:privacy-utility-knn-mti}
\end{figure}

There is one special case in Figure~\ref{fig:trade-off-qnli-mt} that \bert performs slightly worse than \gpt and the baselines when $\epsilon$ is small under MTI attacks. This is mainly due to the fact that BERT used in MTI attack fail to recognize the sub-word tokens in GPT-based \gpt and InferDPT, and word-level tokens in GloVe-based SANTEXT and CUSTEXT. This also explains the high privacy scores (over 70\%) even when $\epsilon=20$. Such vocabulary mismatch actually hinders the attack performance. Therefore, we mainly evaluate KNN attacks in the following experiments.

\subsubsection{Open-ended Text Generation Tasks}\label{sec:exp-open-gen}
Compared to text classification tasks, text generation is more sensitive to the perturbation.
When some key information in the text is disturbed, the generated text may have poor coherence and alignment to the original prompt.
We thereby follow InferDPT~\cite{privinfer-23} to use a lightweight local model (Qwen2-1.5B-Instruct), denoted as \textit{extraction} model $\mathcal{M}_{e}$, to refine the perturbed response generated by the server.
When receiving the noisy generation $\hat{y}$ from $\mathcal{S}$, $\mathcal{C}$ uses the \textit{extraction} model to de-noise the response as $y' \leftarrow \mathcal{M}_{e}(x, \hat{y})$.
Additionally, we denote the expected response (i.e., non-private) as $y$.
In the following experiments, we vary the perturbation mechanisms, while utilizing the same extraction model (default temperature of 0.5). 
The instructions used to generate noisy response and extract response are provided in Appendix~\ref{append:prompt}.
We use \textit{coherence} and \textit{alignment} to measure the text generation utility. Both metrics are computed using cosine similarity that goes as: 
\begin{equation}
    CS(s_1, s_2) = \frac{Emb(s_1) \cdot Emb(s_2)}{|Emb(s_1)| \cdot |Emb(s_2)|}
\end{equation}
where the sentence embedding $Emb(s)$ is computed using the method in SimCSE~\cite{gao2021simcse}. \textbf{Coherence} measures the cosine similarity between original prompt $x$ and the extracted response $y'$ as $CS(x, y')$, and \textbf{Alignment} measures the similarity between the extracted response $y'$ and expected response $y$ as $CS(y, y')$.

\begin{figure}[h]
    \centering
    \setlength{\abovecaptionskip}{3pt}
    \setlength{\belowcaptionskip}{0pt}
    \begin{subfigure}[b]{0.49\linewidth}
        \includegraphics[width=\linewidth]{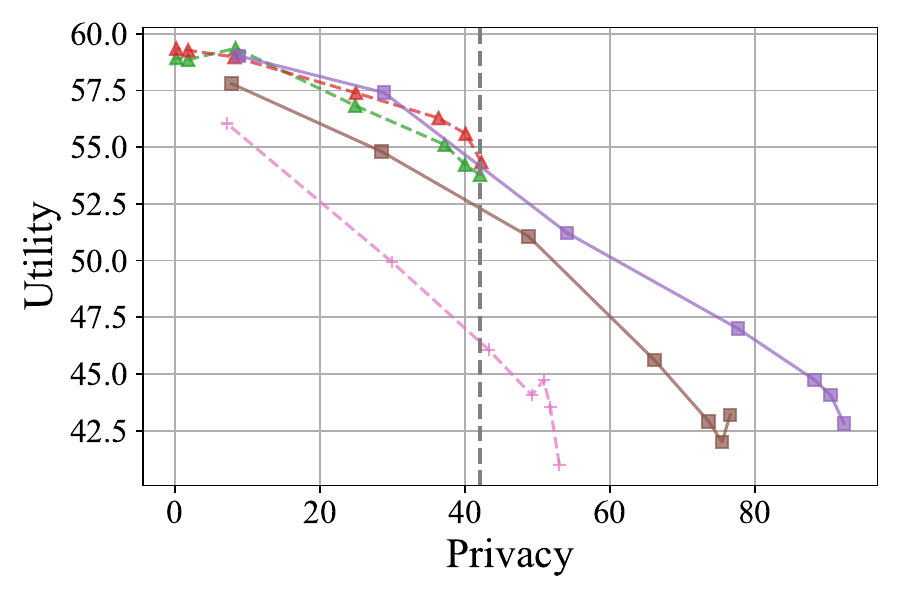}
        \caption{KNN attack vs. Coherence.}
        \label{fig:wiki-knn-coherence}
    \end{subfigure}
    \hfill
    \begin{subfigure}[b]{0.49\linewidth}
        \includegraphics[width=\linewidth]{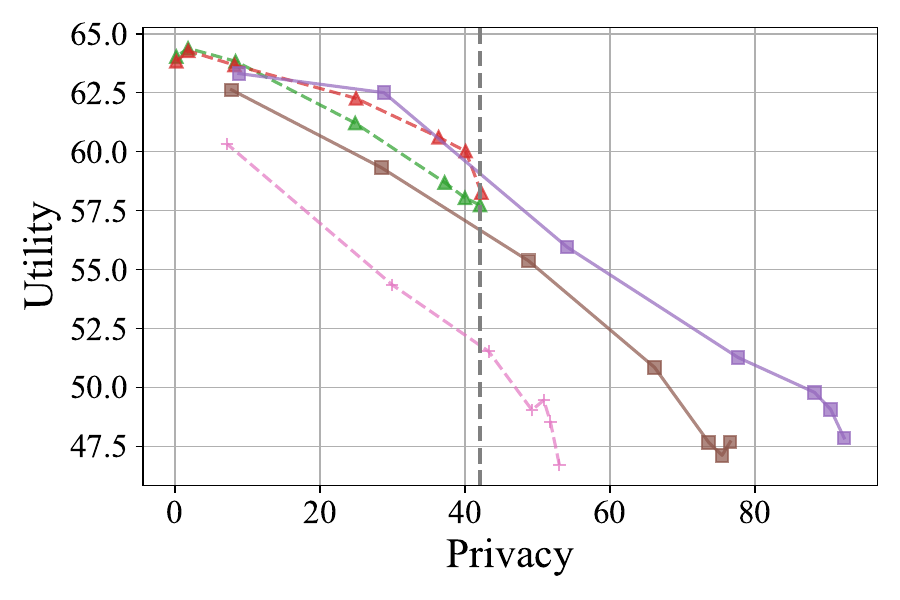}
        \caption{KNN attack vs. Alignment.}
        \label{fig:wiki-knn-alignment}
    \end{subfigure}
    \includegraphics[width=\linewidth]{pic/legend.pdf}
    \caption{Privacy-utility trade-offs in terms of KNN attack by varying $\epsilon \in [1, 20]$ on Wikitext-103-v1 dataset. Privacy is measured by privacy scores under empirical attacks.}
    \label{fig:wiki-privacy-utility}
\end{figure}

Notably, we fail to evaluate SANTEXT because it requires a vocabulary size of approximately 221,642 over Wikitext, which causes OOM error when calculating sampling probabilities. As shown in Figure~\ref{fig:wiki-privacy-utility}, \name significantly outperforms InferDPT and achieves similar utility to CUSTEXT when $\epsilon$ is large. Whereas, CUSTEXT fails to provide reasonable empirical privacy. Even with $\epsilon=1$, it achieves a privacy score of only around 40\%.
The results exhibit the superior performance of \name.
The detailed utility numbers are deferred to Appendix~\ref{append:text-gen-utility}.

\subsection{Influence of Privacy Budget}\label{sec:influence-budget}
In this section, we evaluate each method in terms of formal privacy, with $\epsilon \in [1,20]$. Notably, we examine the effect of varying $\epsilon$ on each method. We do not attempt to align the privacy levels across these methods, as they provide distinct forms of \textit{actual} formal privacy, which could introduce ambiguities.
For instance, the $\epsilon$-metric-LDP-based SANTEXT effectively adheres to $\epsilon'$-DP, where $\epsilon' = \epsilon \cdot d_{max}$, with $d_{max} \sim 14.86$ (as derived from GloVe embeddings, shown in Table~\ref{tab:vocab-size}).
In InferDPT, the random adjacency list is generated using Laplace noise with a default budget of approximately $9$, resulting in an actual privacy budget of $\epsilon' \sim \epsilon + 9$.
Similarly, our method operates under $\epsilon' = \epsilon + \ln(\underset{i, j}{\mathrm{max}}\frac{|b_i|}{|b_j|}) \sim \epsilon + 8$.
An exception is CUSTEXT, which employs a static adjacency list of default size 20, making it hard to quantify the actual privacy budget.


\subsubsection{Utility Evaluation}\label{sec:exp-utility}
We compute the \textbf{Rouge-L F1 score} between the original and perturbed prompts on SST-2 dataset, which serves as an indicator of sentence-level semantic similarity. We also provide some perturbation examples in Appendix~\ref{append:perturbation}.

\begin{table}[h]
    \centering
    \setlength{\abovecaptionskip}{3pt}
    \setlength{\belowcaptionskip}{0pt}
    \caption{Sentence-level similarity of various methods.}
    \scalebox{0.7}{
    \begin{tabular}{@{}cc|ccccccc@{}}
\toprule
\multicolumn{2}{c|}{\multirow{2}{*}{\textbf{Mechanism}}} & \multicolumn{7}{c}{\textbf{Rouge-L (F1)} $\uparrow$}                                                                                                                                                                                               \\ \cmidrule(l){3-9} 
\multicolumn{2}{c|}{}                                    & \multicolumn{1}{c|}{$\epsilon=1$} & \multicolumn{1}{c|}{$\epsilon=2$} & \multicolumn{1}{c|}{$\epsilon=3$} & \multicolumn{1}{c|}{$\epsilon=6$} & \multicolumn{1}{c|}{$\epsilon=10$} & \multicolumn{1}{c|}{$\epsilon=14$} & $\epsilon=20$ \\ \midrule
\multicolumn{2}{c|}{SANTEXT}                             & \multicolumn{1}{c|}{0.87}         & \multicolumn{1}{c|}{13.43}        & \multicolumn{1}{c|}{71.31}        & \multicolumn{1}{c|}{99.36}        & \multicolumn{1}{c|}{99.45}         & \multicolumn{1}{c|}{99.45}         & 99.45         \\
\multicolumn{2}{c|}{SANTEXT+}                            & \multicolumn{1}{c|}{52.56}        & \multicolumn{1}{c|}{57.77}        & \multicolumn{1}{c|}{72.98}        & \multicolumn{1}{c|}{76.37}        & \multicolumn{1}{c|}{77.46}         & \multicolumn{1}{c|}{77.66}         & 77.56         \\ \midrule
\multicolumn{2}{c|}{CUSTEXT}                             & \multicolumn{1}{c|}{14.50}        & \multicolumn{1}{c|}{17.87}        & \multicolumn{1}{c|}{22.74}        & \multicolumn{1}{c|}{47.27}        & \multicolumn{1}{c|}{81.44}         & \multicolumn{1}{c|}{95.54}         & 99.48         \\
\multicolumn{2}{c|}{CUSTEXT+}                            & \multicolumn{1}{c|}{50.66}        & \multicolumn{1}{c|}{52.90}        & \multicolumn{1}{c|}{55.47}        & \multicolumn{1}{c|}{69.41}        & \multicolumn{1}{c|}{88.46}         & \multicolumn{1}{c|}{97.18}         & 99.71         \\ \midrule
\multicolumn{2}{c|}{InferDPT}                            & \multicolumn{1}{c|}{13.00}        & \multicolumn{1}{c|}{14.09}        & \multicolumn{1}{c|}{14.92}        & \multicolumn{1}{c|}{16.48}        & \multicolumn{1}{c|}{23.20}         & \multicolumn{1}{c|}{38.68}         & 68.11         \\ \midrule
\multicolumn{1}{c|}{\multirow{2}{*}{$\name$}}   & Bert   & \multicolumn{1}{c|}{38.38}        & \multicolumn{1}{c|}{39.20}        & \multicolumn{1}{c|}{40.86}        & \multicolumn{1}{c|}{46.85}        & \multicolumn{1}{c|}{60.22}         & \multicolumn{1}{c|}{76.49}         & 92.03         \\ \cmidrule(l){2-9} 
\multicolumn{1}{c|}{}                           & GPT2   & \multicolumn{1}{c|}{37.60}        & \multicolumn{1}{c|}{38.19}        & \multicolumn{1}{c|}{40.08}        & \multicolumn{1}{c|}{44.55}        & \multicolumn{1}{c|}{56.46}         & \multicolumn{1}{c|}{73.46}         & 90.65         \\ \bottomrule
\end{tabular}
}
    \label{tab:rougeL-perturb}
\end{table}

The results are presented in Table~\ref{tab:rougeL-perturb}. In general, the scores of all methods increase with the privacy budget $\epsilon$. Among them, SANTEXT+ and CUSTEXT+ achieve high scores even when $\epsilon$ is small.
This is because in SANTEXT+, a proportion of tokens (top $w\%$ most frequent tokens, even not stopwords) are treated as non-sensitive. While in CUSTEXT, each word has a pre-defined token adjacency list, which results in a high probability that a word will either remain unchanged or be perturbed to a synonym\footnote{The proportion of synonyms exceeds 40\% even when $\epsilon=2$, as shown in Table III of the paper~\cite{privinfer-23}.}, yielding higher similarity. 
This also accounts for their weak privacy defense against attacks.
Regarding SANTEXT, InferDPT and \name, \name achieves the best performance. 
Specifically, when we align the actual $\epsilon\sim 14$, corresponding to SANTEXT ($\epsilon\sim1$), InferDPT and \name ($\epsilon \sim 6$), the score of \name is significantly larger than the other two methods.

\subsubsection{Privacy Evaluation}\label{sec:exp-privacy}
We evaluate the empirical privacy using \textbf{effective mapping set size $S_t$}. This is calculated by counting the number of distinct tokens mapped from $t$ after perturbation as $S_t = \{t'| t' = \mathcal{R}(t)\}$. Intuitively, the mapping set should encompass the entire vocabulary. A larger $S_t$ implies greater entropy in the mapping distribution, thereby indicating higher empirical privacy. 
We also provide two other metrics: \textit{retention ratio $N_t$} and \textit{defense against attacks} in Appendix~\ref{append:retentio-ratio}$\sim$\ref{append:defense-score} due to page limitation.



To evaluate $S_t$, we run perturbation methods 1,000 times to each token in the vocabulary on SST-2 dataset with $\epsilon \in [1, 20]$. 
The results are illustrated in Figure~\ref{fig:sst2-mapping}. 
For the metric-LDP-based SANTEXT, $S_t$ decreases rapidly as $\epsilon$ increases, dropping to approximately 100 when $\epsilon = 3$. In contrast, CUSTEXT maintains $S_t$ at around 20 in most cases, even when $\epsilon = 1$. This consistency stems from CUSTEXT's design, which restricts the sampling space for each token to a fixed size (20 by default). 
However, this approach also implies that under a KNN attack with Top-20 accuracy, CUSTEXT's privacy score would be nearly zero.

\begin{figure}[h]
    \centering
    \captionsetup[subfigure]{skip=0pt}
    \setlength{\abovecaptionskip}{3pt}
    \setlength{\belowcaptionskip}{0pt}
    \begin{subfigure}[b]{0.3\linewidth}
        \includegraphics[width=\linewidth]{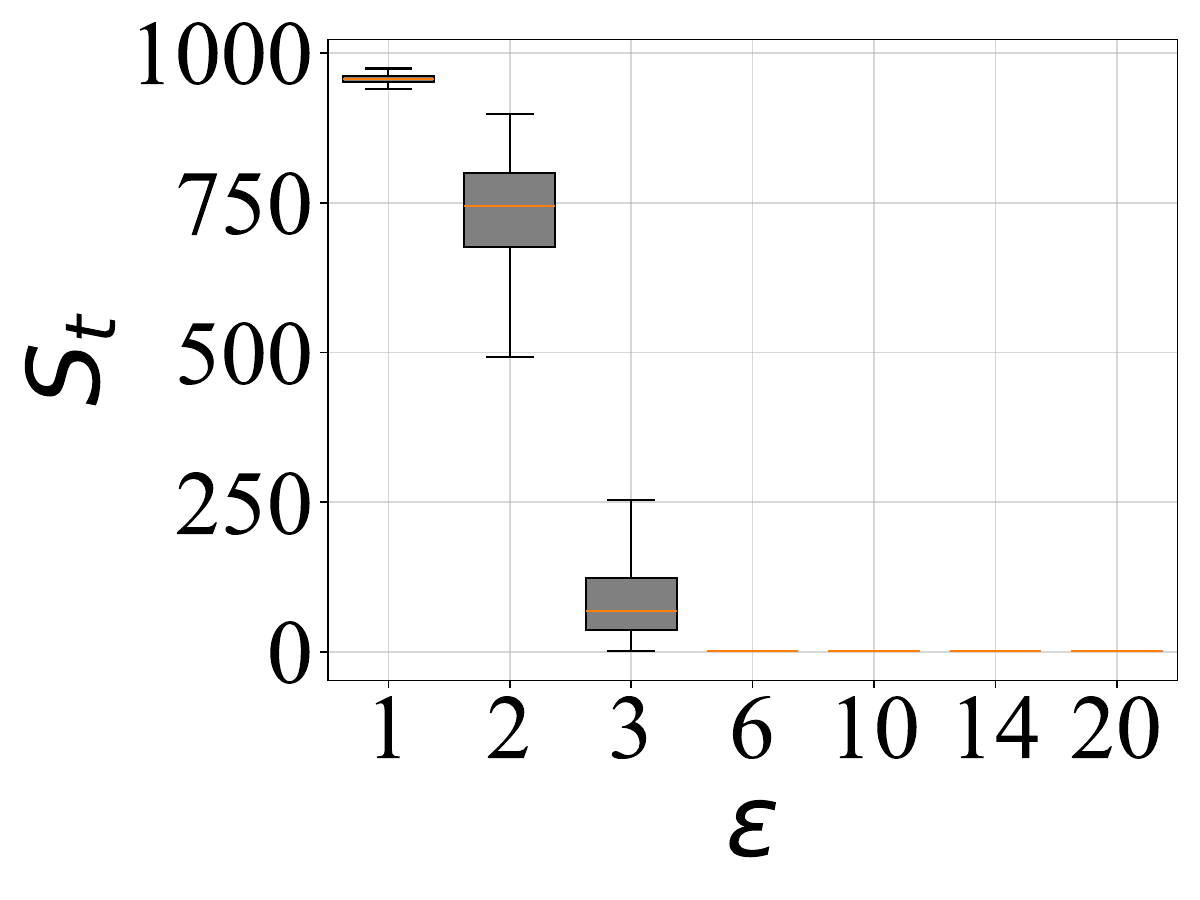}
        \caption{SANTEXT}
        \label{fig:st-san}
    \end{subfigure}
    \begin{subfigure}[b]{0.3\linewidth}
        \includegraphics[width=\linewidth]{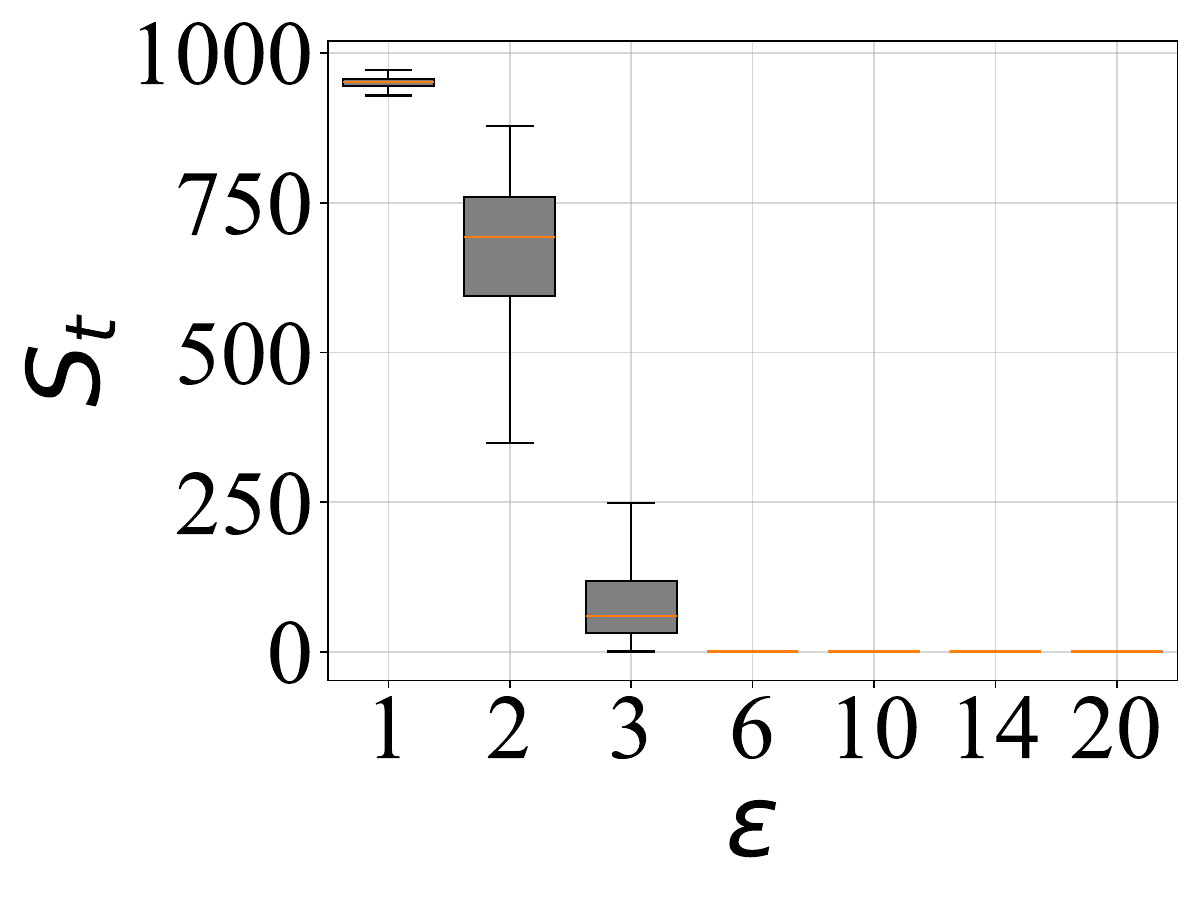}
        \caption{SANTEXT+}
        \label{fig:st-san+}
    \end{subfigure}
    \\
    \begin{subfigure}[b]{0.3\linewidth}
        \includegraphics[width=\linewidth]{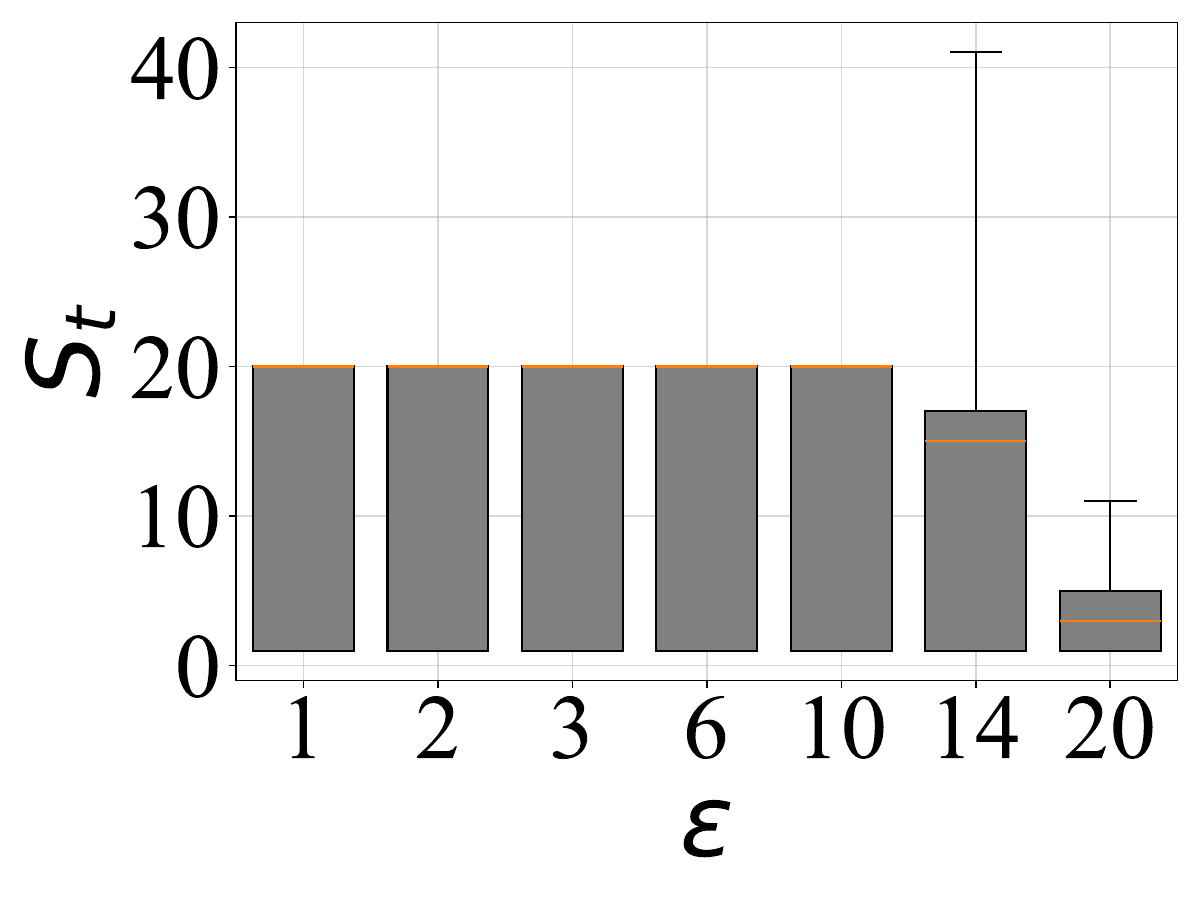}
        \caption{CUSTEXT}
        \label{fig:st-cus}
    \end{subfigure}
    \begin{subfigure}[b]{0.3\linewidth}
        \includegraphics[width=\linewidth]{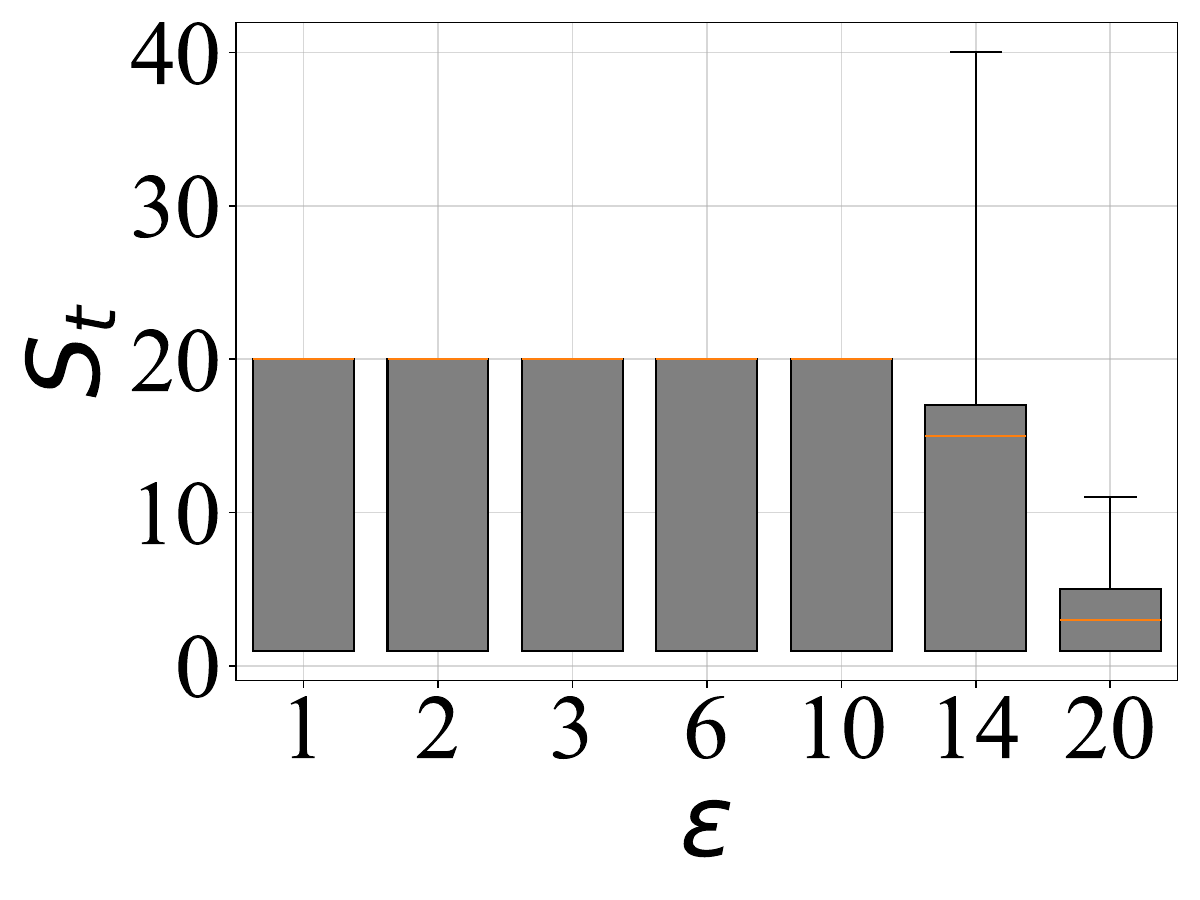}
        \caption{CUSTEXT+}
        \label{fig:st-cus+}
    \end{subfigure}
    \\
    \begin{subfigure}[b]{0.3\linewidth}
        \includegraphics[width=\linewidth]{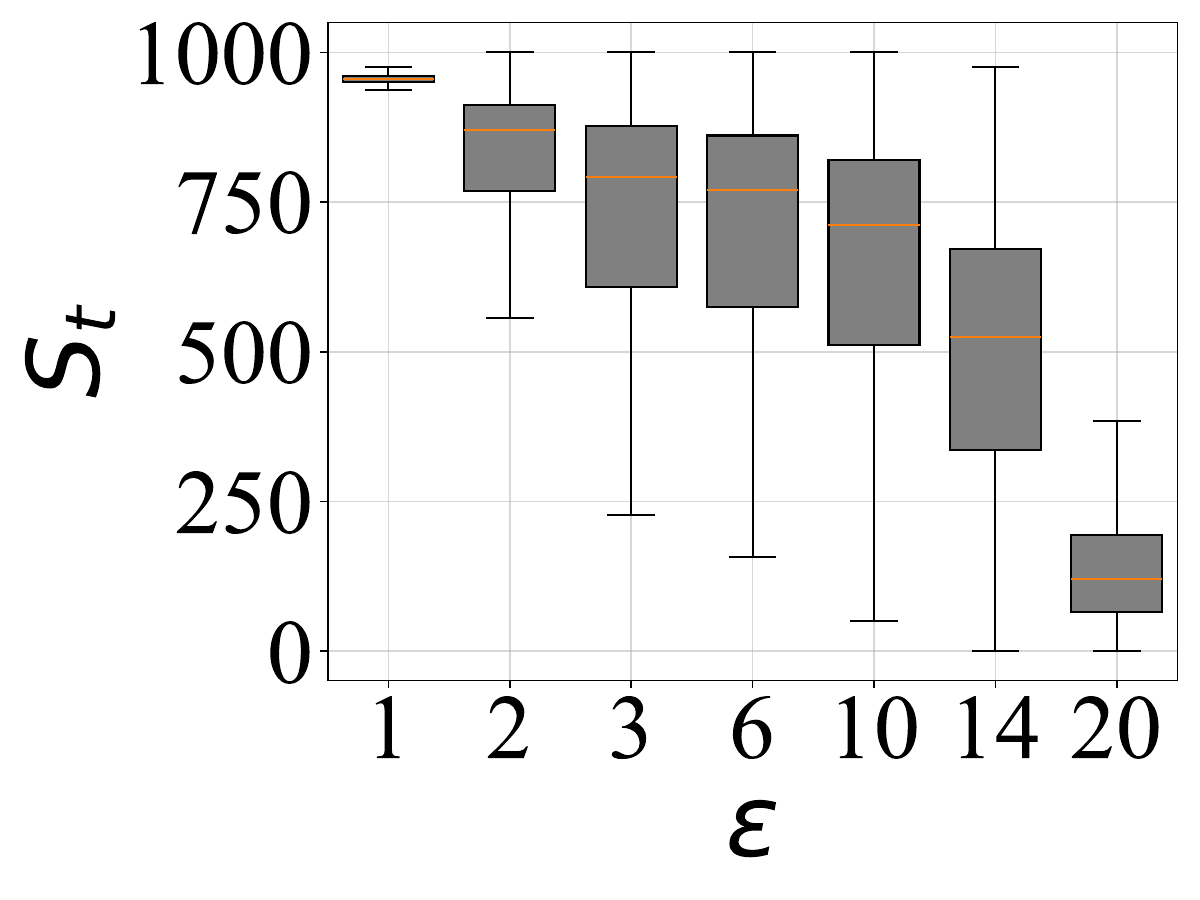}
        \caption{InferDPT}
        \label{fig:st-inferdpt}
    \end{subfigure}
    \begin{subfigure}[b]{0.3\linewidth}
        \includegraphics[width=\linewidth]{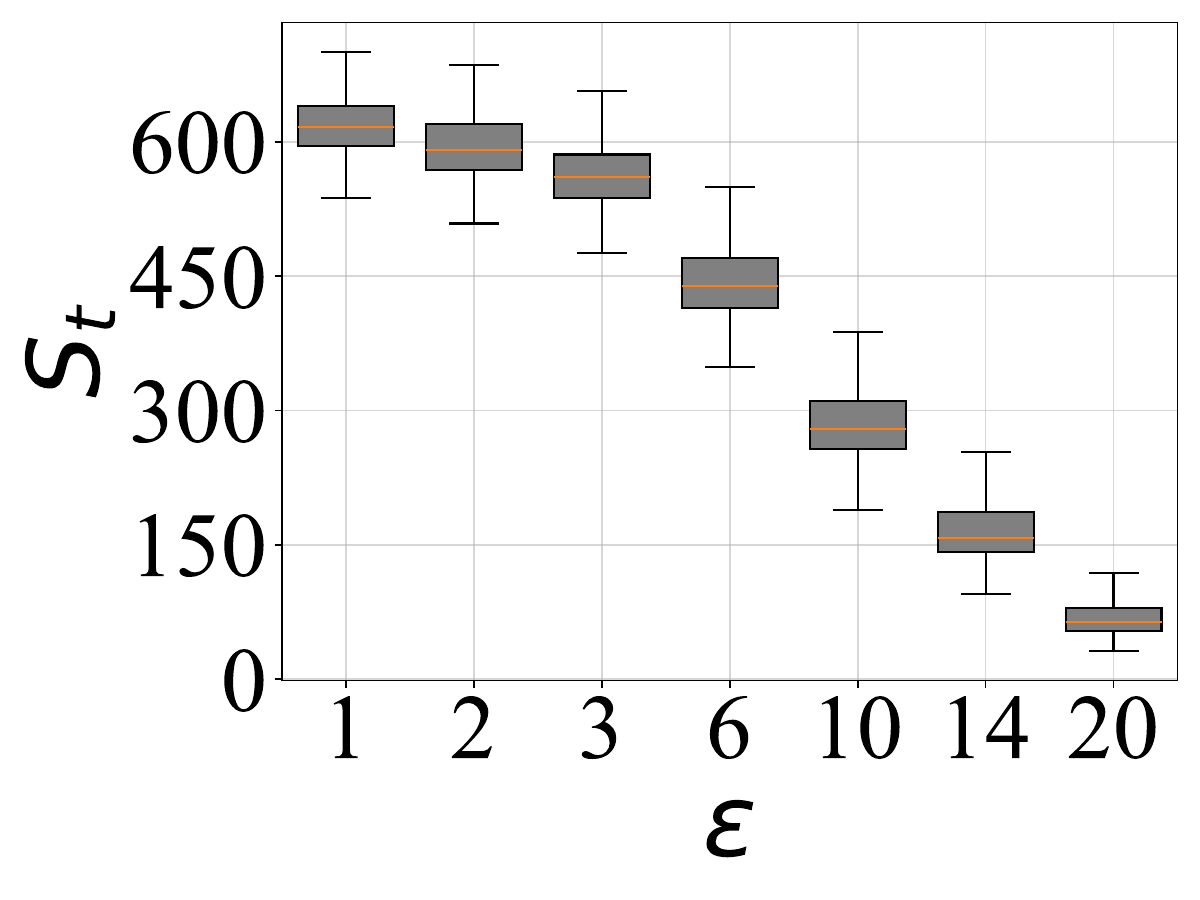}
        \caption{\name (BERT)}
        \label{fig:st-cape-bert}
    \end{subfigure}
    \begin{subfigure}[b]{0.3\linewidth}
        \includegraphics[width=\linewidth]{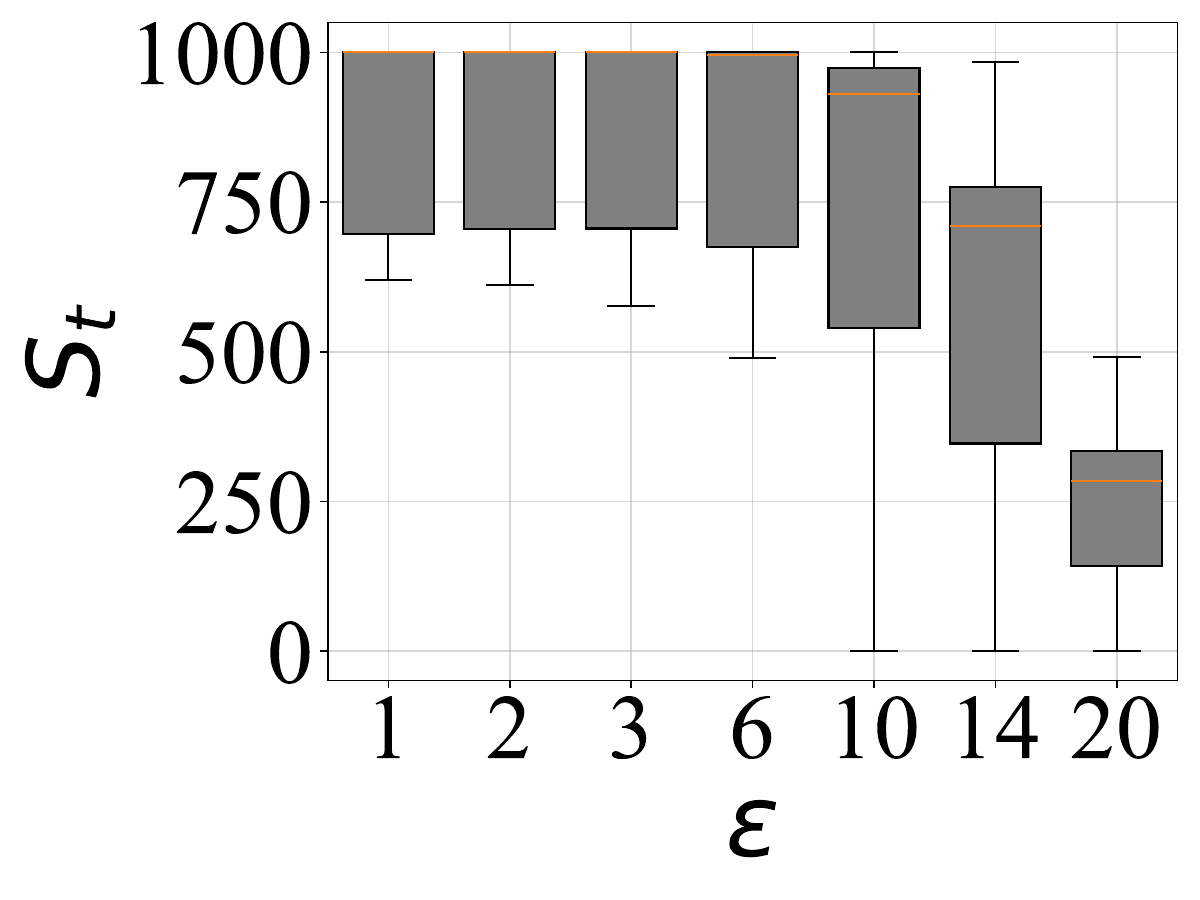}
        \caption{\name (GPT2)}
        \label{fig:st-cape-gpt2}
    \end{subfigure}
    \caption{Mapping set size $S_t$ with $\epsilon \in [1, 20]$ on SST-2.}
    \label{fig:sst2-mapping}
\end{figure}

For InferDPT and \name, both methods exhibit a proportional decrease in $S_t$ as $\epsilon$ increases due to their use of EM over the entire vocabulary. Owing to our bucketing strategy, \name exhibits lower entropy than InferDPT.
Notably, in \name, the BERT variant outperforms the GPT-2 variant by producing a smaller and smoother $S_t$ distribution. This can be attributed to BERT’s use of a denser embedding space. Moreover, BERT is better at capturing both preceding and following context. This distinction also explains why \name (BERT) surpasses \name (GPT-2) in the above experiments.

\subsection{Ablation Studies}\label{sec:exp-extensive}
We additionally evaluate \name under various parameter configurations on SST-2 dataset. From now on, we adopt the Bert-based \name for its superiority.
We here vary utility importance factor and bucket number to perform an extensive comparison. Additionally, we evaluate the feasibility of using \textbf{model distillation} on client-side device model in Appendix~\ref{append:ab-distil}.
We choose KNN attack as the privacy metric and evaluate the classification accuracy on SST-2 dataset. 

\subsubsection{Varying Importance Factor}
To evaluate the effect of incorporating contextual logits, we run experiments on varying logit importance factor $\lambda_L \in [0.2, 0.5, 1.0, 1.2, 1.5]$, with fixed distance importance factor $\lambda_D$ as 1.0 in the hybrid utility function. 

\begin{figure}[h]
    \centering
    \setlength{\abovecaptionskip}{3pt}
    \setlength{\belowcaptionskip}{0pt}
    \begin{subfigure}[b]{0.49\linewidth}
        \includegraphics[width=\linewidth]{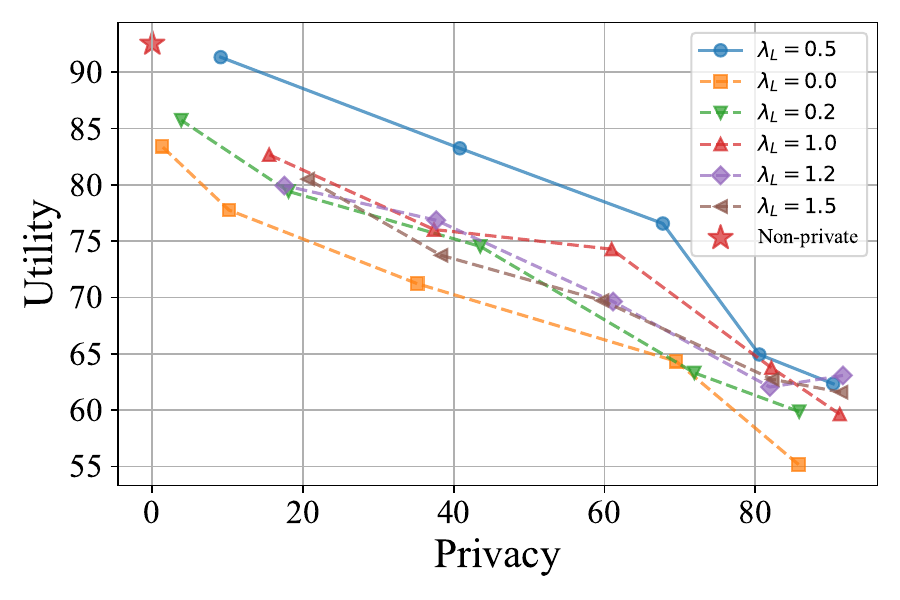}
        \caption{$\lambda_L \in [0, 1.5]$}
        \label{fig:ab-lambdaL}
    \end{subfigure}
    \begin{subfigure}[b]{0.49\linewidth}
        \includegraphics[width=\linewidth]{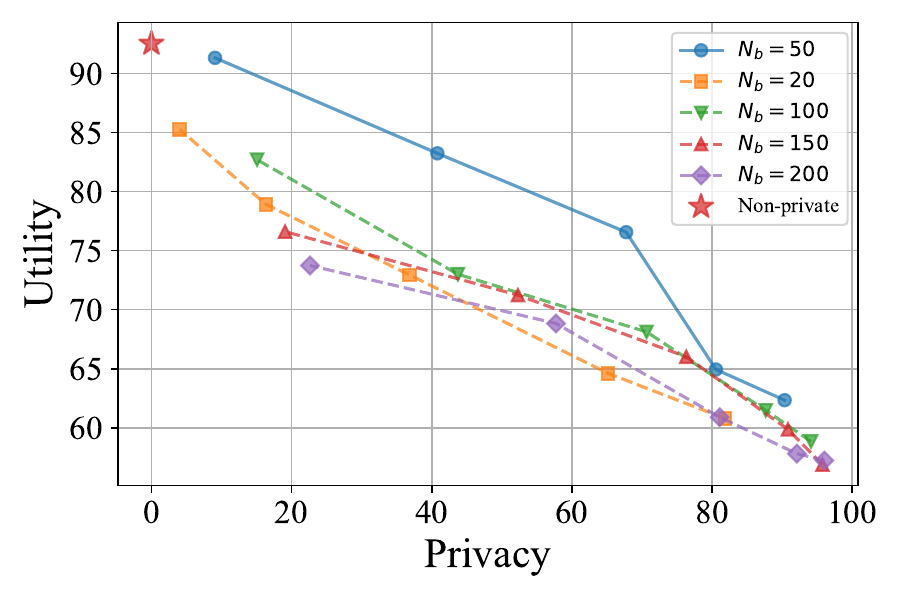}
        \caption{$N_b \in [20, 200]$}
        \label{fig:ab-bucket}
    \end{subfigure}
    \caption{Ablation results under various parameter configurations with $\epsilon \in [1,20]$.}
    \label{fig:ablation-bld}
\end{figure}

As shown in Figure~\ref{fig:ab-lambdaL}, a utility function that relies solely on token embedding distance (i.e., $\lambda_L = 0.0$) demonstrates the poorest performance, highlighting the importance of incorporating contextual logits for designing effective utility functions. However, assigning excessive weight to logits (e.g., $\lambda_L = 1.5$) results in diminished performance. This occurs because a token that better fits the context may not always be semantically similar to the original token.

\subsubsection{Varying Bucket Number}
We use bucketing to handle the dilemma of large sampling space. To test the effect of bucket number $N_b$, we conduct experiments on varying $N_b \in \{20, 50, 100, 150, 200\}$. 

The experiment results are illustrated in Figure~\ref{fig:ab-bucket}. 
Notably, $N_b=50$ achieves the best performance, while the two extremes, $N_b \in \{20, 200\}$, yield the worst results. As $N_b$ increases, resulting in a larger sampling space, the empirical privacy level improves. For instance, when $\epsilon=1.0$, the privacy score for $N_b=200$ is approximately 10\% higher than that for $N_b=50$. However, this comes at the cost of utility, which declines rapidly. This is because larger $N_b$ makes it more likely to select a bucket with low utility, causing significant utility degradation. Balancing the privacy-utility trade-off, $N_b=50$ emerges as an optimal choice. 
\section{Conclusion}\label{sec:conclusion}
In this paper, we propose a context-aware prompt perturbation mechanism to strike a better balance in privacy-utility trade-off in LLM inference services. By perturbing user prompts using differential privacy, we provide quantifiable privacy protection, with moderate utility degradation. In the future, we plan to explore more fine-grained privacy mechanisms tailored to different groups of tokens, such as application-specific semantic categories or user-specific personalization scenarios.


\section*{Impact Statement}
This paper presents work whose goal is to advance the field of Machine Learning. Our work aims to address critical privacy concerns surrounding user prompts in widely adopted LLM-based inference services. To safeguard sensitive user information during inference, we leverage differential privacy, which provides quantifiable privacy protection controlled by a privacy budget. However, we also acknowledge the inevitable utility degradation caused by the introduction of random noise to perturb the original prompts. A key contribution of this work lies in the development of a novel hybrid utility function that integrates contextual information with token embedding distances, alongside a bucketized sampling algorithm designed specifically for the NLP domain's large sampling space. These advancements have the potential to make private LLM inference more practical and scalable for real-world applications.

\bibliography{citation}
\bibliographystyle{icml2025}

\newpage
\appendix
\onecolumn

\section{Vocabulary Configuration}~\label{append:vocab}
The vocabulary configuration for the baselines and \name are provided in Table~\ref{tab:vocab-size}.
Notably, SANTEXT leverages a subset of GloVe, specifically tailored to the dataset's vocabulary, excluding tokens that do not appear in the dataset.
CUSTEXT similarly trims the GloVe vocabulary to align with a predefined vocabulary subset~\tablefootnote{Limited vocabulary: \url{https://github.com/nmrksic/counter-fitting/blob/master/linguistic\_constraints/vocabulary.txt}}. InferDPT also prunes the cl100k\_base to initial 11,000 English tokens as the vocabulary. Specifically, the embeddings in CUSTEXT are normalized, resulting in smaller Euclidean distances.

\begin{table*}[h]
    \centering
    \caption{Vocabulary configurations. $\mathcal{V}_{sample}$ denotes the actual sampling space. $K$ and $\hat{K}$ denote the size of (random) adjacency list in CUSTEXT and InferDPT, respectively.}
    \scalebox{0.8}{
    \begin{tabular}{@{}c|ccc|ccc|ccc@{}}
\toprule
\multirow{2}{*}{\textbf{Methods}} & \multicolumn{3}{c|}{\textbf{\begin{tabular}[c]{@{}c@{}}GloVe\\ $(d=300)$\end{tabular}}}                                                                                      & \multicolumn{3}{c|}{\textbf{\begin{tabular}[c]{@{}c@{}}Bert\\ $(d=768)$\end{tabular}}} & \multicolumn{3}{c}{\textbf{\begin{tabular}[c]{@{}c@{}}GPT\\ $(d \in \{768, 1536\})$\end{tabular}}} \\ \cmidrule(l){2-10} 
                                  & $|\mathcal{V}|$ & $d_{max}$                                             & $|\mathcal{V}_{sample}|$                                                                           & $|\mathcal{V}|$            & $d_{max}$            & $|\mathcal{V}_{sample}|$           & $|\mathcal{V}|$                 & $d_{max}$               & $|\mathcal{V}_{sample}|$               \\ \midrule
SANTEXT                           & 2,196,017       & 14.86 & \begin{tabular}[c]{@{}c@{}}14,283 (SST-2)\\ 87,751 (QNLI)\\ 221,642 (Wikitext-103-v1)\end{tabular} & -                          & -                    & -                                  & -                               & -                       & -                                      \\ \midrule
CUSTEXT                           & 76,855          & 1.66                                                  & $K$                                                                                                & -                          & -                    & -                                  & -                               & -                       & -                                      \\ \midrule
InferDPT                          & -               & -                                                     & -                                                                                                  & -                          & -                    & -                                  & 100,256 (11,000)                & 1.44                    & $\hat{K}$                              \\ \midrule
$\name$                           & -               & -                                                     & -                                                                                                  & 30,522                     & 2.89                 & 30,522                             & 50,257                          & 7.58                    & 50,257                                 \\ \bottomrule
\end{tabular}
}
    \label{tab:vocab-size}
\end{table*}

\section{Detailed Experiment Results}

\subsection{Utility for text classification tasks}~\label{append:classification-utility}
Table~\ref{tab:utility} presents the utility of different methods, measured by inference accuracy of text classification on SST-2 and QNLI datasets.
Notably, as mentioned in Section~\ref{sec:influence-budget}, we directly use the definition of privacy budget $\epsilon$ in their papers without alignment to our work in case of ambiguities. We here focus on the influence of $\epsilon$.

\begin{table}[h]
    \centering
    \caption{Utility comparison of various perturbation mechanisms at similar privacy levels on SST-2 and QNLI datasets.}
    \scalebox{0.7}{
    \begin{tabular}{@{}cc|ccccccc|ccccccc@{}}
\toprule
\multicolumn{2}{c|}{\multirow{2}{*}{\textbf{Methods}}} & \multicolumn{7}{c|}{\textbf{SST-2}}                                                                       & \multicolumn{7}{c}{\textbf{QNLI}}                                                                         \\ \cmidrule(l){3-16} 
\multicolumn{2}{c|}{}                                  & $\epsilon=1$ & $\epsilon=2$ & $\epsilon=3$ & $\epsilon=6$ & $\epsilon=10$ & $\epsilon=14$ & $\epsilon=20$ & $\epsilon=1$ & $\epsilon=2$ & $\epsilon=3$ & $\epsilon=6$ & $\epsilon=10$ & $\epsilon=14$ & $\epsilon=20$ \\ \midrule
\multicolumn{2}{c|}{Random}                            & \multicolumn{7}{c|}{51.72}                                                                                & \multicolumn{7}{c}{50.78}                                                                                 \\ \midrule
\multicolumn{2}{c|}{SANTEXT}                           & 52.29        & 52.52        & 76.15        & 91.06        & 92.09         & 91.17         & 90.14         & 50.98        & 50.91        & 66.68        & 76.90        & 76.55         & 76.55         & 76.66         \\
\multicolumn{2}{c|}{SANTEXT+}                          & 64.22        & 67.20        & 77.52        & 79.47        & 81.63         & 82.17         & 83.27         & 58.72        & 60.44        & 66.26        & 68.74        & 69.52         & 71.13         & 71.59         \\ \midrule
\multicolumn{2}{c|}{CUSTEXT}                           & 62.50        & 63.53        & 65.60        & 72.01        & 82.21         & 91.51         & 92.20         & 61.71        & 61.61        & 63.23        & 67.07        & 73.81         & 76.11         & 76.57         \\
\multicolumn{2}{c|}{CUSTEXT+}                          & 71.56        & 72.13        & 73.05        & 82.56        & 86.19         & 91.74         & 92.43         & 65.22        & 66.48        & 67.38        & 69.69        & 76.00         & 76.66         & 77.01         \\ \midrule
\multicolumn{2}{c|}{InferDPT}                          & 54.19        & 55.52        & 56.54        & 58.38        & 56.42         & 60.44         & 79.70         & 52.27        & 53.33        & 54.19        & 56.13        & 57.67         & 62.80         & 72.18         \\ \midrule
\multicolumn{1}{c|}{\multirow{2}{*}{$\name$}}  & Bert  & 58.14        & 60.01        & 62.33        & 64.93        & 76.56         & 83.24         & 91.33         & 52.10        & 51.77        & 52.17        & 53.74        & 59.76         & 67.93         & 75.84         \\ \cmidrule(l){2-16} 
\multicolumn{1}{c|}{}                          & GPT2  & 59.63        & 60.89        & 61.27        & 63.99        & 77.81         & 80.67         & 90.25         & 51.66        & 51.86        & 52.59        & 52.66        & 59.45         & 67.09         & 73.62         \\ \midrule
\multicolumn{2}{c|}{Non-private}                       & \multicolumn{7}{c|}{92.54}                                                                                & \multicolumn{7}{c}{76.88}                                                                                 \\ \bottomrule
\end{tabular}
}
    \label{tab:utility}
\end{table}

\subsection{Utility for open-ended text generation tasks}~\label{append:text-gen-utility}
The detailed utility numbers for open-ended text generation tasks are illustrated in Table~\ref{tab:text-gen-utility}. We measure the utility of the noisy generation from two aspects: \textit{coherence} to the original prompt, and \textit{alignment} to the expected generation.
\begin{table*}[h]
    \centering
    \caption{Coherence \& Similarity on Open-ended Text Generation.}
    \scalebox{0.7}{
\begin{tabular}{@{}cc|ccccccc|ccccccc@{}}
\toprule
\multicolumn{2}{c|}{\multirow{2}{*}{\textbf{Methods}}} & \multicolumn{7}{c|}{\textbf{Coherence $\uparrow$}}                                                        & \multicolumn{7}{c}{\textbf{Alignment $\uparrow$}}                                                         \\ \cmidrule(l){3-16} 
\multicolumn{2}{c|}{}                                  & $\epsilon=1$ & $\epsilon=2$ & $\epsilon=3$ & $\epsilon=6$ & $\epsilon=10$ & $\epsilon=14$ & $\epsilon=20$ & $\epsilon=1$ & $\epsilon=2$ & $\epsilon=3$ & $\epsilon=6$ & $\epsilon=10$ & $\epsilon=14$ & $\epsilon=20$ \\ \midrule
\multicolumn{2}{c|}{CUSTEXT}                           & 53.78        & 54.23        & 55.11        & 56.83        & 59.36         & 58.85         & 58.93         & 57.73        & 58.04        & 58.69        & 61.21        & 63.83         & 64.38         & 64.03         \\
\multicolumn{2}{c|}{CUSTEXT+}                          & 54.34        & 55.59        & 56.28        & 57.39        & 58.97         & 59.27         & 59.34         & 58.26        & 60.02        & 60.61        & 62.26        & 63.67         & 64.27         & 63.83         \\ \midrule
\multicolumn{2}{c|}{InferDPT}                          & 41.00        & 43.53        & 44.73        & 44.07        & 46.07         & 49.93         & 56.04         & 46.72        & 48.55        & 49.46        & 49.04        & 51.53         & 54.35         & 60.33         \\ \midrule
\multicolumn{1}{c|}{\multirow{2}{*}{$\name$}}  & Bert  & 42.82        & 44.07        & 44.74        & 47.00        & 51.22         & 55.40         & 58.02         & 47.86        & 49.07        & 49.79        & 51.27        & 55.95         & 59.50         & 62.31         \\ \cmidrule(l){2-16} 
\multicolumn{1}{c|}{}                          & GPT   & 43.20        & 42.01        & 42.91        & 45.61        & 51.07         & 54.81         & 57.80         & 47.69        & 47.12        & 47.67        & 50.85        & 55.37         & 59.32         & 62.62         \\ \midrule
\multicolumn{2}{c|}{Non-private}                       & \multicolumn{7}{c|}{59.78}                                                                                & \multicolumn{7}{c}{65.05}                                                                                 \\ \bottomrule
\end{tabular}
}
    \label{tab:text-gen-utility}
\end{table*}

\subsection{Privacy: Retention Ratio}~\label{append:retentio-ratio}
Retention ratio $N_t$ is calculated by counting the frequency of retention of token $t$ after the perturbation.
This is a straightforward metric for evaluating privacy protection. A higher $N_t$ indicates that more tokens remain unperturbed, reflecting a lower level of privacy protection. As the privacy budget $\epsilon$ increases, all methods exhibit higher $N_t$. Among these methods, \name consistently achieves the lowest $N_t$. 
SANTEXT demonstrates a much faster increase in $N_t$ due to its usage of metric-LDP. Additionally, since SANTEXT+ classifies a portion of the most frequent tokens as non-sensitive, and InferDPT operates with a truncated vocabulary of size 11,000, both methods exhibit high $N_t$ even when $\epsilon=1$. 
CUSTEXT also produces higher $N_t$ due to its limited sampling space for each token. Under a fixed $\epsilon$, the $N_t$ of CUSTEXT is nearly double that of \name.

\begin{figure}[h]
    \centering
    \includegraphics[width=0.8\linewidth]{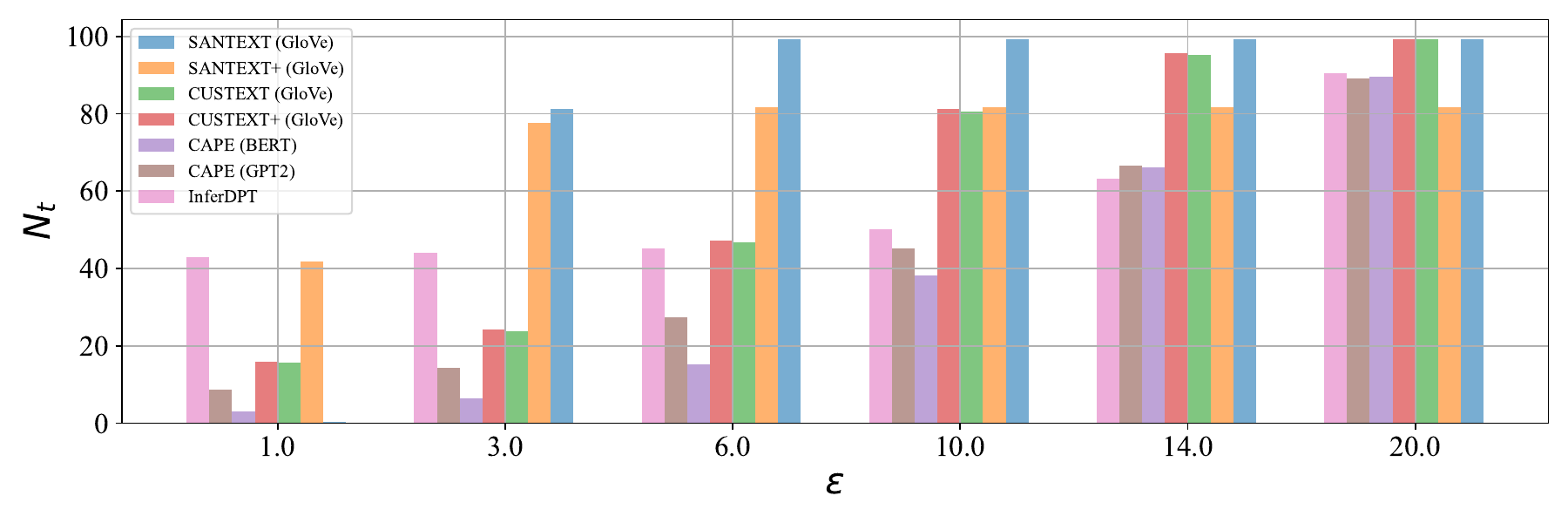}
    \caption{Retention ratio $N_t$ by varying the privacy parameter $\epsilon \in [1, 20]$ on the SST-2 dataset.}
        \label{fig:sst2-nt}
\end{figure}

\subsection{Privacy: Defense scores against privacy attacks}~\label{append:defense-score}
We conduct some existing attacks to empirically demonstrate the privacy protection comparison with varying privacy budgets. As introduced in Section~\ref{sec:experiments}, we calculate the attack success rate $asr$ on sensitive tokens and the privacy score as $1 - asr$.

\begin{figure}[h]
    \centering
    \begin{subfigure}[b]{0.24\linewidth}
        \includegraphics[width=\linewidth]{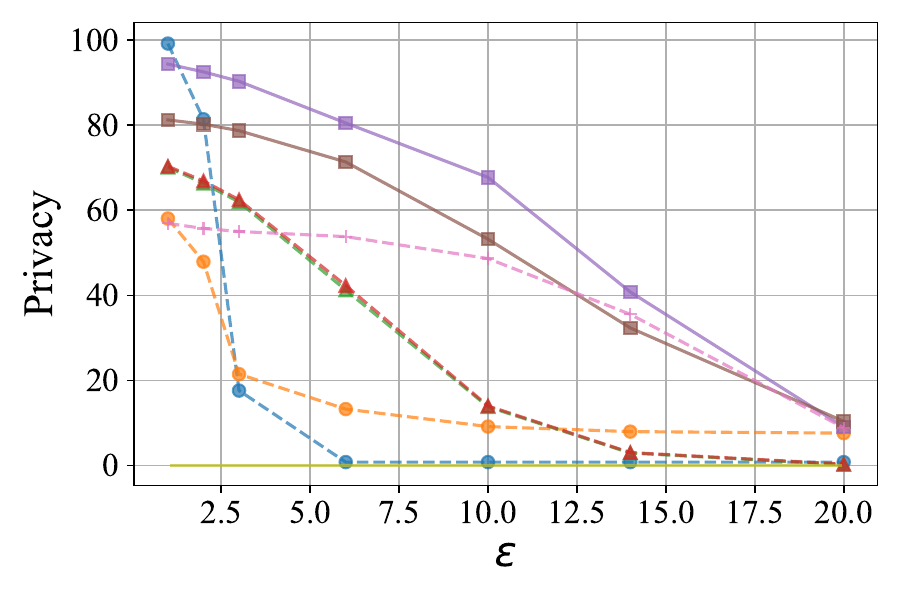}
        \caption{KNN attacks on SST-2.}
        \label{fig:sst2-knn}
    \end{subfigure}
    \hfill
    \begin{subfigure}[b]{0.24\linewidth}
        \includegraphics[width=\linewidth]{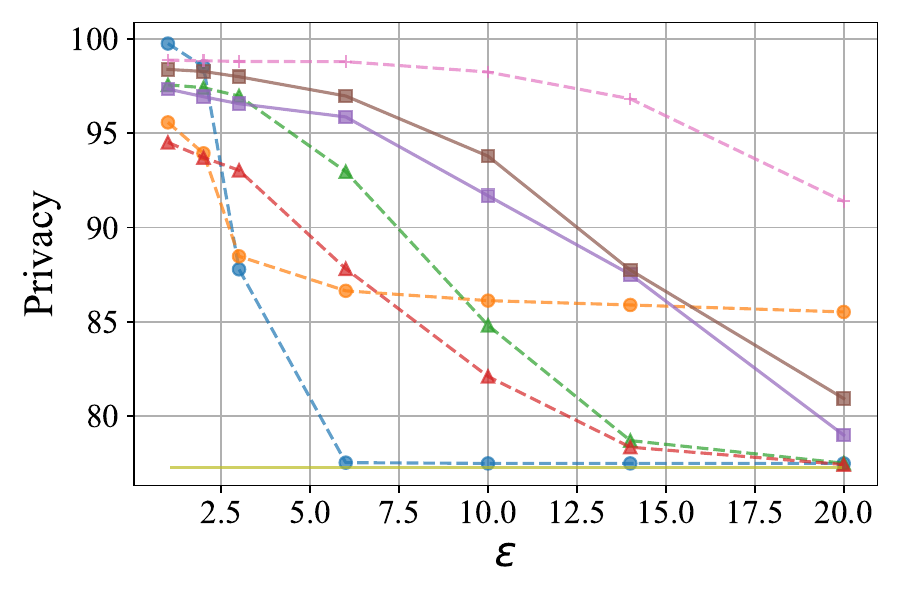}
        \caption{MTI attacks on SST-2.}
        \label{fig:sst2-mt}
    \end{subfigure}
    \begin{subfigure}[b]{0.24\linewidth}
        \includegraphics[width=\linewidth]{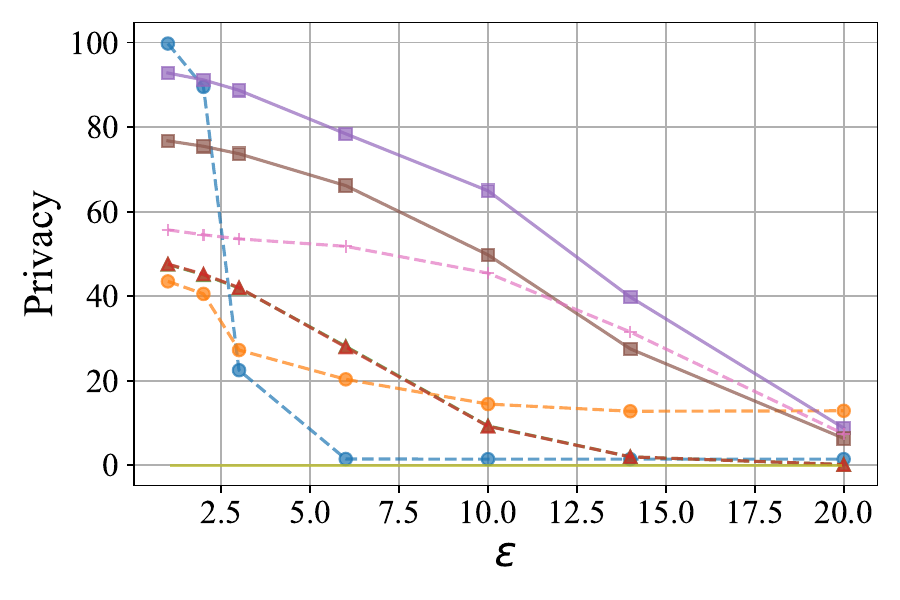}
        \caption{KNN attacks on QNLI.}
        \label{fig:qnli-knn}
    \end{subfigure}
    \hfill
    \begin{subfigure}[b]{0.24\linewidth}
        \includegraphics[width=\linewidth]{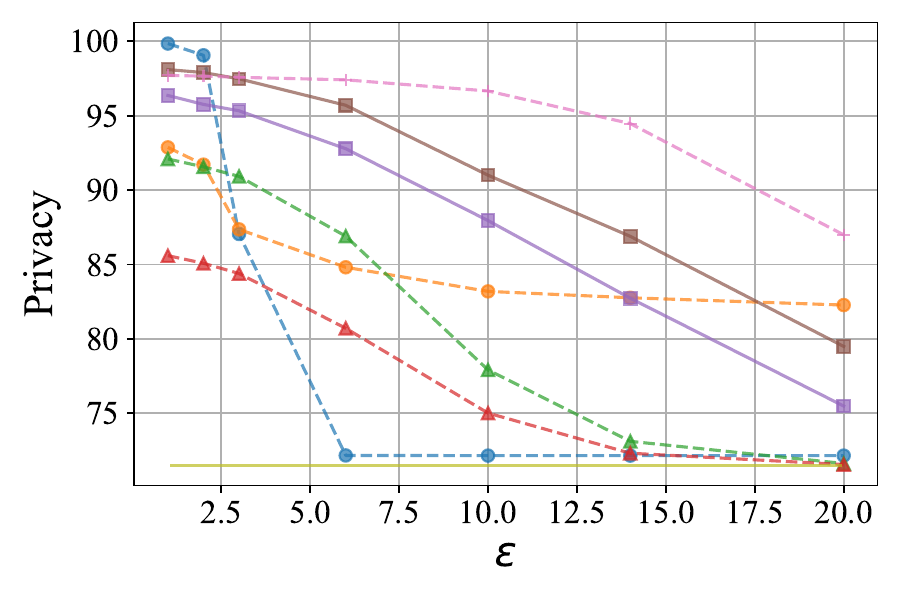}
        \caption{MTI attacks on QNLI.}
        \label{fig:qnli-mt}
    \end{subfigure}
    \includegraphics[width=0.7\linewidth]{pic/legend.pdf}
    \caption{Privacy evaluation by varying the privacy parameter $\epsilon \in [1, 20]$.}
    \label{fig:sst2-privacy}
\end{figure}

\subsection{Efficiency Evaluation}\label{append:efficiency}
We evaluate the efficiency of various methods by measuring their runtime cost and memory consumption. A batch size of 2 is used, representing two simultaneous client-side queries. This is reasonable given the lack of high concurrency requirements on the client. Each experiment is repeated five times to calculate the average runtime.
The total runtime is divided into two phases: 1) \textbf{Setup} phase, which pre-computes token embedding distances, constructs adjacency lists and calculates sampling probabilities; 2) \textbf{Perturbation} phase, which performs the perturbation on a input user prompt.

As shown in Table~\ref{tab:efficiency}, \name ensures prompt privacy with minimal time overhead. While \name consumes more resources due to the use of a local model, its average runtime per input prompt remains under 0.2 seconds. SANTEXT and CUSTEXT exhibit the lowest runtime costs, as they precompute adjacency lists and sampling probabilities for each token. However, this comes at the expense of higher setup phase runtimes, particularly for CUSTEXT, where adjacency list construction significantly increases overhead.
The GPU memory consumption of \name is approximately $1\sim1.5$ GB. In Appendix~\ref{append:ab-distil}, we further explore the potential of model distillation to reduce this memory requirement.

\begin{table}[h]
    \centering
    \caption{Efficiency evaluation by measuring the average runtime cost in each stage.}
    \setlength{\tabcolsep}{25pt}
    \scalebox{0.8}{
    \begin{tabular}{@{}cc|cc|c@{}}
\toprule
\multicolumn{2}{c|}{\multirow{2}{*}{\textbf{Methods}}} & \multicolumn{2}{c|}{\textbf{Runtime}}        & \textbf{Memory} \\ \cmidrule(l){3-5} 
\multicolumn{2}{c|}{}                                  & \multicolumn{1}{c|}{Setup}    & Perturbation & Perturbation    \\ \midrule
\multicolumn{2}{c|}{SANTEXT}                           & \multicolumn{1}{c|}{325.14s}  & 0.003s       & 0.81G           \\ \midrule
\multicolumn{2}{c|}{CUSTEXT}                           & \multicolumn{1}{c|}{2929.75s} & 0.002s       & 0.16G           \\ \midrule
\multicolumn{2}{c|}{InferDPT}                          & \multicolumn{1}{c|}{28.66s}   & 0.05s        & 1.03G           \\ \midrule
\multicolumn{1}{c|}{\multirow{2}{*}{$\name$}}  & Bert  & \multicolumn{1}{c|}{37.91s}   & 0.15s        & 1.37G           \\ \cmidrule(l){2-5} 
\multicolumn{1}{c|}{}                          & GPT2  & \multicolumn{1}{c|}{98.25s}   & 0.10s        & 1.12G           \\ \bottomrule
\end{tabular}
}
    \label{tab:efficiency}
\end{table}

\subsection{Ablation: Distilled Local Model}~\label{append:ab-distil}

In \name, a client-side device model is used to obtain the contextual information to enhance the utility. The parameter size for the Bert-base and GPT2-base models evaluated above is about 400 MB. We note that for those resource-limited devices, we need to further decrease the resource consumption. Model distillation~\cite{distil-bert} provides a promising direction to lower the model size. To verify its feasibility in \name, we evaluated the performance on the widely-used distilbert model~\footnote{\url{https://huggingface.co/distilbert/distilbert-base-uncased}.}, of which the size is reduced by about 40\%.

\begin{figure}[h]
    \centering
    \begin{subfigure}[b]{0.45\linewidth}
        \includegraphics[width=\linewidth]{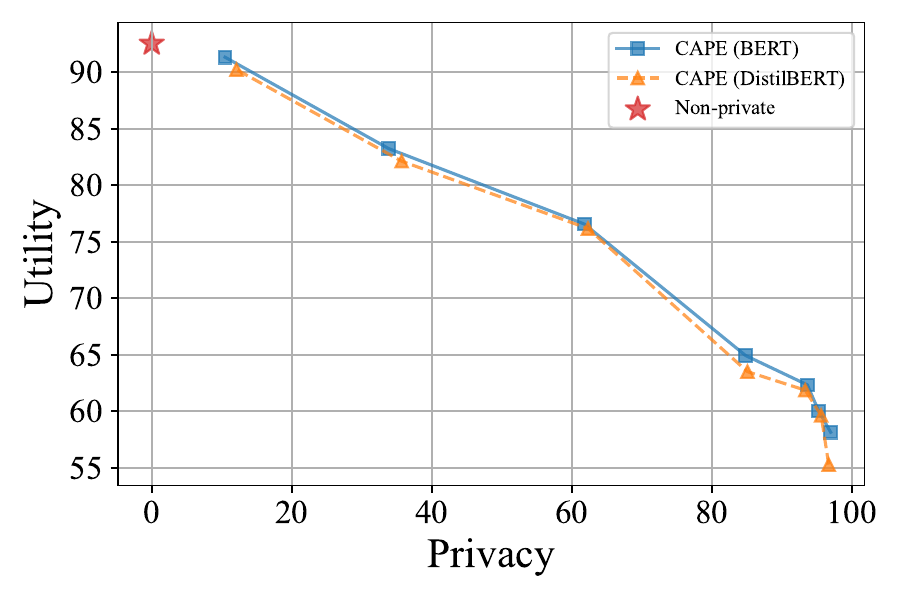}
        \caption{KNN attack with Top-1 accuracy.}
        \label{fig:ab-top1}
    \end{subfigure}
    \hfill
    \begin{subfigure}[b]{0.45\linewidth}
        \includegraphics[width=\linewidth]{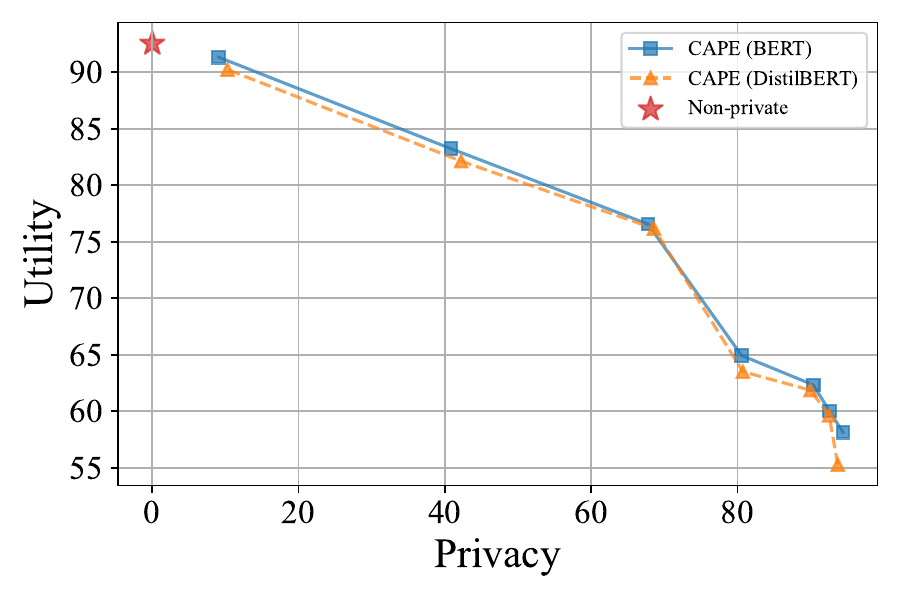}
        \caption{KNN attack with Top-10 accuracy}
        \label{fig:ab-top10}
    \end{subfigure}
    \caption{Privacy-utility trade-offs in terms of success rates of KNN attacks vs. accuracy rates by varying the privacy parameter $\epsilon \in [1, 20]$ on the SST-2 dataset.}
    \label{fig:ablation-distil}
\end{figure}

We evaluate the classification accuracy and KNN attack accuracy on the SST-2 dataset using both the bert-base and distilbert models. As illustrated in Figure~\ref{fig:ablation-distil}, distilbert generally achieves performance comparable to the original bert model. Specifically, its classification accuracy is at most $\sim 1\%$ lower than that of bert under the same empirical privacy level. Detailed results in Table~\ref{tab:ablation-distil} indicate that the distilled model exhibits slightly lower accuracy and stronger privacy defense on a fixed formal priavcy level (i.e., $\epsilon$). This can be attributed to the reduced expressiveness of distilbert embeddings compared to the original model.

\begin{table*}[h]
    \centering
    \caption{Privacy-utility trade-offs in terms of success rates of KNN attacks vs. accuracy rates w/wo model distillation.}
    \scalebox{0.65}{
\begin{tabular}{@{}cc|ccccccc|ccccccc@{}}
\toprule
\multicolumn{2}{c|}{\multirow{2}{*}{\textbf{Methods}}}     & \multicolumn{7}{c|}{\textbf{Accuracy $\uparrow$}}                                                         & \multicolumn{7}{c}{\textbf{Privacy (Top-10) $\uparrow$}}                                                  \\ \cmidrule(l){3-16} 
\multicolumn{2}{c|}{}                                      & $\epsilon=1$ & $\epsilon=2$ & $\epsilon=3$ & $\epsilon=6$ & $\epsilon=10$ & $\epsilon=14$ & $\epsilon=20$ & $\epsilon=1$ & $\epsilon=2$ & $\epsilon=3$ & $\epsilon=6$ & $\epsilon=10$ & $\epsilon=14$ & $\epsilon=20$ \\ \midrule
\multicolumn{1}{c|}{\multirow{2}{*}{$\name$}} & Bert       & 58.14        & 60.01        & 62.33        & 64.93        & 76.56         & 83.24         & 91.33         & 94.40        & 92.56        & 90.34        & 80.54        & 67.74         & 40.80         & 9.07          \\ \cmidrule(l){2-16} 
\multicolumn{1}{c|}{}                         & DistilBert & 55.28        & 59.63        & 61.86        & 63.50        & 76.17         & 82.13         & 90.24         & 93.69        & 92.50        & 89.96        & 80.72        & 68.57         & 42.21         & 10.27         \\ \midrule
\multicolumn{2}{c|}{Non-private}                           & \multicolumn{7}{c|}{92.54}                                                                                & \multicolumn{7}{c}{100.0}                                                                                 \\ \bottomrule
\end{tabular}
}
    \label{tab:ablation-distil}
\end{table*}

\subsection{Histogram of logits and token embedding distance}~\label{append:hist}
We here run some statistics on the example prompt: `This is a good \underline{book}.' using GPT-2 model. We compute the logits for \underline{book} with `This is a good [MASK].' as the context. We compute the Euclidean distance between \underline{book} and other tokens in the vocabulary $\mathcal{V}$. The histogram of the logits and distance are illustrated in Figure~\ref{fig:hist}.
\begin{figure}[h]
    \centering
    \begin{subfigure}[b]{0.45\linewidth}
        \includegraphics[width=\linewidth]{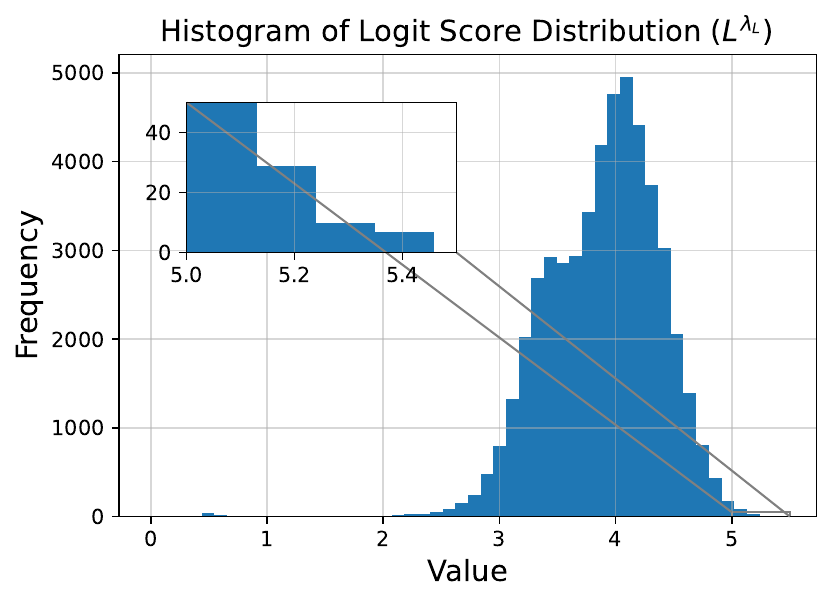}
        \caption{Logit $L^{\lambda_L}$}
        \label{fig:log-score}
    \end{subfigure}
    \hfill
    \begin{subfigure}[b]{0.45\linewidth}
        \includegraphics[width=\linewidth]{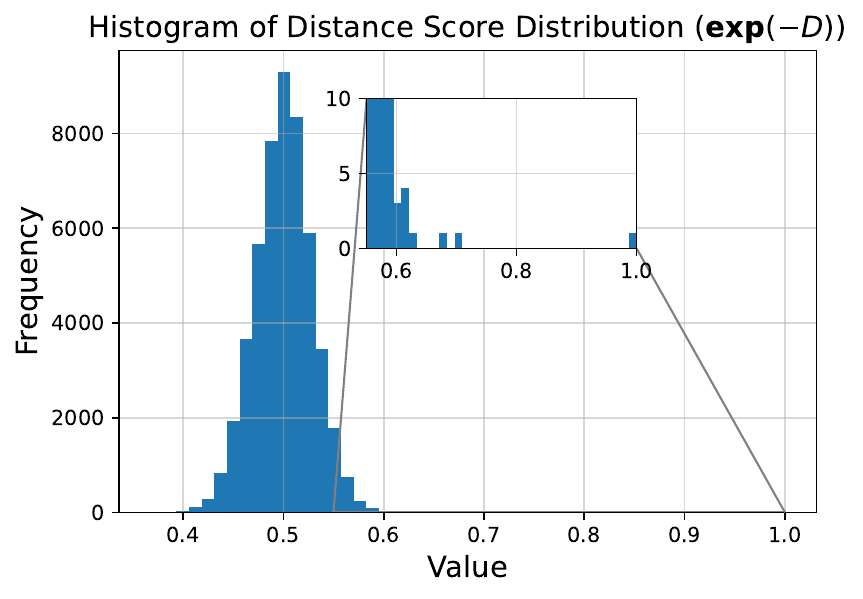}
        \caption{Distance $\mathbf{exp}(-D)^{\lambda_D}$}
        \label{fig:dist-score}
    \end{subfigure}
    \caption{Histogram of different distance metrics for `This is a good \underline{book}.'. $\lambda_L = 0.5, \lambda_D=1.0$.}
    \label{fig:hist}
\end{figure}

\section{Proofs of Theorems}\label{append:proof}
\begin{proof}[Proof of Theorem~\ref{theo:bem}]
    Consider the original token $t \in \mathcal{V}$, another token $t' \in \mathcal{V}\ \backslash\ \{t\}$ and a possible output token $y \in \mathcal{V}$.
    We use $B(t) = b_i (i\in [N_b])$ to denote the $i$-th bucket that token $t$ belongs to, $N_b$ is the bucket number. Additionally, we use $u_b(t, \cdot)$ to denote the bucket utility for $B(t)$ and $|B(t)|$ to denote the cardinality of bucket $B(t)$.

According to the sampling probabilities in Equation~\ref{eq:bucket-prob}, we have:
\begin{align*}
    &\frac{\mathbb{P}[\mathcal{R}(t)=y]}{\mathbb{P}[\mathcal{R}(t')=y]} \\
    =&(\frac{\mathbf{exp}(\frac{\epsilon}{2\triangle}u_b(t, y))}{\sum_{j=1}^{|\mathcal{V}|}\mathbf{exp}(\frac{\epsilon}{2\triangle}u_b(t, y_j))} \cdot \frac{1}{|B(y)|}) / (\frac{\mathbf{exp}(\frac{\epsilon}{2\triangle}u_b(t', y))}{\sum_{j=1}^{|\mathcal{V}|}\mathbf{exp}(\frac{\epsilon}{2\triangle}u_b(t', y_j))} \cdot \frac{1}{|B'(y)|}) \\
    =&\frac{\mathbf{exp}(\frac{\epsilon}{2\triangle}u_b(t, y))}{\mathbf{exp}(\frac{\epsilon}{2\triangle}u_b(t', y))} \cdot \frac{\sum_{j=1}^{|\mathcal{V}|}\mathbf{exp}(\frac{\epsilon}{2\triangle}u_b(t', y_j))}{\sum_{j=1}^{|\mathcal{V}|}\mathbf{exp}(\frac{\epsilon}{2\triangle}u_b(t, y_j))} \cdot \frac{|B'(y)|}{|B(y)|}\\
\end{align*}
Let $\underset{i, j}{\mathrm{max}}\frac{|b_i|}{|b_j|} = \mathbf{exp}(\epsilon')$, we have $\frac{|B'(y)|}{|B(y)|} \leq \mathbf{exp}(\epsilon')$. Consequently, we have:
\begin{align*}
    &\frac{\mathbb{P}[\mathcal{R}(t)=y]}{\mathbb{P}[\mathcal{R}(t')=y]} \\
    \leq &\frac{\mathbf{exp}(\frac{\epsilon}{2\triangle}u_b(t, y))}{\mathbf{exp}(\frac{\epsilon}{2\triangle}u_b(t', y))} \cdot \frac{\sum_{j=1}^{|\mathcal{V}|}\mathbf{exp}(\frac{\epsilon}{2\triangle}u_b(t', y_j))}{\sum_{j=1}^{|\mathcal{V}|}\mathbf{exp}(\frac{\epsilon}{2\triangle}u_b(t, y_j))} \cdot \mathbf{exp}(\epsilon')\\
    =&\mathbf{exp}[\frac{\epsilon}{2}\cdot\frac{(u(t, y) - u(t', y))}{\triangle}] \cdot \frac{\sum_{j=1}^{|\mathcal{V}|}\mathbf{exp}(\frac{\epsilon}{2\triangle}u(t', y_j))}{\sum_{j=1}^{|\mathcal{V}|}\mathbf{exp}(\frac{\epsilon}{2\triangle}u(t, y_j))}\cdot \mathbf{exp}(\epsilon')\\
    \leq&\mathbf{exp}(\frac{\epsilon}{2})\cdot\frac{\sum_{j=1}^{|\mathcal{V}|}\mathbf{exp}(\frac{\epsilon}{2\triangle}u(t', y_j))}{\sum_{j=1}^{|\mathcal{V}|}\mathbf{exp}(\frac{\epsilon}{2\triangle}u(t, y_j))}\cdot \mathbf{exp}(\epsilon')\\
    \leq&\mathbf{exp}(\frac{\epsilon}{2}) \cdot \mathbf{exp}(\frac{\epsilon}{2}) \cdot \mathbf{exp}(\epsilon') \\
    =&\mathbf{exp}(\epsilon + \epsilon')
\end{align*}
The proof shown that our sampling mechanism satisfies ($\epsilon + \epsilon'$)-DP.
\end{proof}

\section{System Prompts}\label{append:prompt}
We provide the system prompts used in the text classification and open-ended text generation tasks in Table~\ref{tab:prompt-detail}.
\begin{table}[h]
\caption{The detailed prompts for text classification and open-ended text generation tasks.}  
\label{tab:prompt-detail}
\centering
\small
\begin{tabularx}{\linewidth}{X}

    \toprule
    \coloredbox{lightpurple}{Text Classification on SST-2:} \\
    You are a helpful assistant. Classify if the sentence is positive or negative sentiment. Just gives the answer in positive, negative.\\
    ----``Content'': \{input\}\\
    \midrule
    
   \coloredbox{lightyellow}{Text Classification on QNLI:}\\
    You are a helpful assistant trained to determine whether a given context sentence contains the information needed to answer the given question. If the context answers the question, respond "Yes". Otherwise, respond "No".\\
    ----``Question'': \{question\}\\
    ----``Context'': \{sentence\}\\
    \midrule
    
  \coloredbox{SoftRed}{Text Generation on Wikitext-103-v1:} \\
    Your task is to extend Prefix Text.\\
    ----``Prefix Text'': \{prompt\}\\
    Provide only your Continuation.\\
    ----``Continuation'':\\
    \midrule
    
   \coloredbox{SoftBlue}{Extraction of noisy response on Wikitext-103-v1:} \\
    Your task is to extend the ``Prefix Text''.\\
    Use the “Perturbed Generation” as your primary writing material for your extension. \\
    Extract coherent and consistent text from the ``Perturbed Generation'' and integrate them into your continuation. \\
    Ensure a seamless alignment with the context established by the ``Prefix Text''.\\
    Provide only your ``Extended Text''.\\
    ----``Prefix Text'': \{prompt\}\\
    ----``Perturbed Generation'': \{noisy\_response\}\\
    ----``Extended Text'':\\ 
    \bottomrule
\end{tabularx}
\end{table}

\section{Perturbation Example}\label{append:perturbation}
We provide some perturbation examples in Table~\ref{tab:perturb_example}. Given input prompts `\textbf{it 's a charming and often affecting journey .}' and `\textbf{it 's slow -- very , very slow .}', we vary the perturbation methods and privacy budgets.

\begin{table}[h]
    \centering
    \caption{Perturbation examples of varying privacy budgets $\epsilon$ for different methods.}
    \scalebox{0.7}{
    \begin{tabular}{@{}|c|c|l|c|@{}}
\toprule
\textbf{Mechanism}        & \textbf{$\epsilon$} & \textbf{\begin{tabular}[c]{@{}l@{}}Original Prompt:\\ \\ it 's a charming and often affecting journey . \\ it 's slow -- very , very slow .\end{tabular}}                                           & \textbf{Rouge-L (F1)} \\ \midrule
\multirow{4}{*}{SANTEXT}  & 1                   & \begin{tabular}[c]{@{}l@{}}photo describes several brilliant nowhere practically partly sorts nietzsche\\ clearly do chewing bolstered female reid see addictive emphasized\end{tabular}            & 0.87                  \\ \cmidrule(l){2-4} 
                          & 2                   & \begin{tabular}[c]{@{}l@{}}nevertheless serviceable 2/3 reaction pretend especially daft journey .\\ sucked addition laziness -- increasingly catastrophic defies unwatchable injuries\end{tabular} & 13.43                 \\ \cmidrule(l){2-4} 
                          & 3                   & \begin{tabular}[c]{@{}l@{}}fan recognized a charming its often affecting journey .\\ yes 's slow -- very , intensity stacks .\end{tabular}                                                          & 71.31                 \\ \cmidrule(l){2-4} 
                          & 6                   & \begin{tabular}[c]{@{}l@{}}it 's a charming and often affecting journey .\\ it 's slow -- very , very slow .\end{tabular}                                                                           & 99.36                 \\ \midrule
\multirow{4}{*}{SANTEXT+} & 1                   & \begin{tabular}[c]{@{}l@{}}it 's a charming and often committed journey .\\ bible 's slow laconic very , very slow solely\end{tabular}                                                              & 52.56                 \\ \cmidrule(l){2-4} 
                          & 2                   & \begin{tabular}[c]{@{}l@{}}discussed 's combined charming banquet often majority encountered precedent\\ it 's slow -- very , very slow .\end{tabular}                                              & 57.77                 \\ \cmidrule(l){2-4} 
                          & 3                   & \begin{tabular}[c]{@{}l@{}}indeed 's leaving charming bothered often operates journey f.\\ it 's slow -- very , very slow .\end{tabular}                                                            & 72.98                 \\ \cmidrule(l){2-4} 
                          & 6                   & \begin{tabular}[c]{@{}l@{}}it wanted puzzled enchanting and often affecting journey yes\\ honestly s unfortunately attract however possibilities very slow .\end{tabular}                           & 76.37                 \\ \midrule
\multirow{4}{*}{CUSTEXT}  & 1                   & \begin{tabular}[c]{@{}l@{}}this 's rather charming all may impair alone .\\ kind 's try -- something , this went .\end{tabular}                                                                     & 14.50                 \\ \cmidrule(l){2-4} 
                          & 2                   & \begin{tabular}[c]{@{}l@{}}as 's rather charming others tend impair alone .\\ kind 's try -- something , this went .\end{tabular}                                                                   & 17.87                 \\ \cmidrule(l){2-4} 
                          & 3                   & \begin{tabular}[c]{@{}l@{}}as 's be charming also often impair future .\\ kind 's try -- sort , this went .\end{tabular}                                                                            & 22.74                 \\ \cmidrule(l){2-4} 
                          & 6                   & \begin{tabular}[c]{@{}l@{}}rather 's a charming also remain impair happiness .\\ every 's slow -- very , is try .\end{tabular}                                                                      & 47.27                 \\ \midrule
\multirow{4}{*}{CUSTEXT+} & 1                   & \begin{tabular}[c]{@{}l@{}}it 's a lovely and exist consequence experiences .\\ it 's runs -- very , very slow .\end{tabular}                                                                       & 50.66                 \\ \cmidrule(l){2-4} 
                          & 2                   & \begin{tabular}[c]{@{}l@{}}it 's a lovely and we consequence experiences .\\ it 's runs -- very , very slow .\end{tabular}                                                                          & 52.90                 \\ \cmidrule(l){2-4} 
                          & 3                   & \begin{tabular}[c]{@{}l@{}}it 's a lovely and we altering experiences .\\ it 's runs -- very , very slow .\end{tabular}                                                                             & 55.47                 \\ \cmidrule(l){2-4} 
                          & 6                   & \begin{tabular}[c]{@{}l@{}}it 's a fairytale and often affecting lifestyle .\\ it 's pace -- very , very slow .\end{tabular}                                                                        & 69.41                 \\ \midrule
\multirow{4}{*}{InferDPT} & 1                   & \begin{tabular}[c]{@{}l@{}}Saw ' axis amar Translate Sie Incorrect invasion sizes . \\ security ' instance exter -- precision , Converter versus .\end{tabular}                                     & 13.00                 \\ \cmidrule(l){2-4} 
                          & 6                   & \begin{tabular}[c]{@{}l@{}}plash ' sha outed Char TE prior affecting Joy .\\ ILL ' ana neglect -- extraordinary , intimate tox .\end{tabular}                                                       & 16.48                 \\ \cmidrule(l){2-4} 
                          & 10                  & \begin{tabular}[c]{@{}l@{}}Kit ' so Adventure charming ing according affecting analysis .\\ Tit ' s l -- verity , functional wrest .\end{tabular}                                                   & 23.20                 \\ \cmidrule(l){2-4} 
                          & 14                  & \begin{tabular}[c]{@{}l@{}}inferior ' Bike a rub and alike affecting journey .\\ learning ' n slow -- very , levant instead .\end{tabular}                                                          & 38.68                 \\ \midrule
\multirow{4}{*}{$\name$}  & 1                   & \begin{tabular}[c]{@{}l@{}}it's a royals andnard popularized progressing. dormitory\\ it's libre - - very, very slower. '\end{tabular}                                                              & 38.38                 \\ \cmidrule(l){2-4} 
                          & 6                   & \begin{tabular}[c]{@{}l@{}}it's axon and quietlyately journey. will\\ it's - - very, very civilized. of\end{tabular}                                                                                & 46.85                 \\ \cmidrule(l){2-4} 
                          & 10                  & \begin{tabular}[c]{@{}l@{}}it's a charming and greatly intimidation her.\\ it's slower - - very, very busy.\end{tabular}                                                                            & 60.22                 \\ \cmidrule(l){2-4} 
                          & 14                  & \begin{tabular}[c]{@{}l@{}}it's a charming and highly painful journey.\\ it's slow - - very, very slow.\end{tabular}                                                                                & 76.49                 \\ \midrule
\multirow{4}{*}{$\name$}  & 1                   & \begin{tabular}[c]{@{}l@{}}it's aivably and typicallyEStream screened. Christensen\\ it'sDragonMagazine after very, very 8.Keefe\end{tabular}                                                       & 37.60                 \\ \cmidrule(l){2-4} 
                          & 6                   & \begin{tabular}[c]{@{}l@{}}it's a really and her angry smart. all\\ it's poor downed very, veryadium. roommate\end{tabular}                                                                         & 44.55                 \\ \cmidrule(l){2-4} 
                          & 10                  & \begin{tabular}[c]{@{}l@{}}it's a telling and often affecting ride.\\ it's a -- very, very PM. where\end{tabular}                                                                                   & 56.46                 \\ \cmidrule(l){2-4} 
                          & 14                  & \begin{tabular}[c]{@{}l@{}}it's a w and often affecting journey.\\ it's slow office very, very put. This\end{tabular}                                                                               & 73.46                 \\ \bottomrule
\end{tabular}
}
    \label{tab:perturb_example}
\end{table}

\paragraph{Toy Example.}
Consider the example prompt: `it 's slow -- very , very slow .'. The step-by-step perturbation process is as follows.
\begin{enumerate}
    \item Initialize an empty token set as $T \leftarrow \phi$.
    \item Tokenize the text prompt using tokenizer and obtain a list of tokens denoted as $L$.
    \item For each token $t$ in the list, calculate its utility score. Take `slow' in `it 's \underline{slow} -- very , very slow .' as an example.
    \begin{enumerate}
        \item Compute the contextual logits for the \textsf{MASKED} token as $L=\mathcal{M}(\text{`it 's [\textsf{MASKED}] -- very , very slow .'})$.
        \item Compute the embedding distance between `slow' and other tokens in the vocabulary as $D = d_{euc}(\text{`slow'}, \mathcal{V})$.
        \item Obtain the final utility score as $u = L^{\lambda_L}\cdot D^{\lambda_D}$.
        \item Bucketize all the tokens in the vocabulary into $N_b$ buckets according to the utility score
        \item Sample a bucket using standard Exponential mechanism.
        \item Uniformly sample a token $\hat{t}$ from the bucket sampled in the previous step.
        \item Concatenate the newly sampled token to $T$ as $T \leftarrow T|\hat{t}$.
    \end{enumerate}
    \item Finally, we get the perturbed prompt $T$.
\end{enumerate}

\end{document}